\documentclass[12pt]{article}

\usepackage{color,amsmath}
\usepackage{times}
\usepackage{graphicx}
\usepackage{multirow}

\usepackage{mhchem}
\usepackage{comment}
\usepackage[hyphens]{url}
\usepackage{hyperref}
\usepackage[hyphenbreaks]{breakurl}

\usepackage{braket}
\usepackage[sort,square,numbers]{natbib}
\usepackage{authblk}

\usepackage{pdfpages}

\usepackage{xr-hyper}
\makeatletter
\newcommand*{\addFileDependency}[1]{
  \typeout{(#1)}
  \@addtofilelist{#1}
  \IfFileExists{#1}{}{\typeout{No file #1.}}
}
\makeatother

\newcounter{lastnote}

\newcommand{\poly}[1]{\text{poly}{#1}}
\newcommand{\rev}[1]{{\textcolor{black}{#1}}}

\author[1]{Seunghoon Lee}
\author[2]{Joonho Lee}
\author[1]{Huanchen Zhai}
\author[3]{Yu Tong}
\author[4]{Alexander M. Dalzell}
\author[5,9]{Ashutosh Kumar}
\author[1]{Phillip Helms}
\author[1]{Johnnie Gray}
\author[1]{Zhi-Hao Cui}
\author[1]{Wen-Yuan Liu}
\author[4,6]{Michael Kastoryano}
\author[7]{Ryan Babbush}
\author[10,4]{John Preskill} 
\author[2]{David R. Reichman}
\author[11]{Earl T. Campbell}
\author[5]{Edward F. Valeev}
\author[3,8]{Lin Lin}
\author[1]{Garnet Kin-Lic Chan \thanks{gkc1000@gmail.com}}

\affil[1]{Division of Chemistry and Chemical Engineering, California Institute of Technology, Pasadena, California 91125, USA}
\affil[2]{Department of Chemistry, Columbia University, New York, New York 10027, USA}
\affil[3]{Department of Mathematics, University of California, Berkeley, California 94720, USA}
\affil[4]{AWS Center for Quantum Computing, Pasadena, California 91125, USA}
\affil[5]{Department of Chemistry, Virginia Tech, Blacksburg, Virginia 24061, USA}
\affil[6]{Amazon Quantum Solutions Lab, Seattle, Washington 98170, USA}
\affil[7]{Google Quantum AI, 340 Main Street, Venice, CA 90291, USA}
\affil[8]{Computational Research Division, Lawrence Berkeley National Laboratory, Berkeley, California 94720, USA}
\affil[9]{Theoretical Division, Los Alamos National Laboratory, Los Alamos, NM 87545, USA}
\affil[10]{Institute for Quantum Information and Matter, California Institute of Technology, Pasadena, California 91125, USA}
\affil[11]{Riverlane, Cambridge, UK}

\date{}

\begin{document} 

\baselineskip24pt

\title{Is there evidence for exponential quantum advantage in quantum chemistry?}
\maketitle

\clearpage

\begin{abstract}
The idea to use quantum mechanical devices to simulate other quantum systems is \rev{commonly} ascribed to Feynman.
Since \rev{the original suggestion, concrete} proposals have appeared for simulating molecular and materials chemistry through quantum computation, as a potential ``killer \rev{application}''\rev{~\cite{reiher2017elucidating,cen, bauer2020quantum,cao2019quantum, mcardle2020quantum}}. 
\rev{Indications of potential} exponential quantum advantage in artificial tasks\rev{~\cite{arute2019quantum,zhong2020quantum, wu2021strong, hangleiter2022computational}} have increased interest in this application, thus, it is critical to understand the basis for potential exponential quantum advantage in quantum chemistry. Here we gather the evidence for this case in the most common task in quantum chemistry, namely, ground-state energy estimation. We conclude that evidence for such an \rev{exponential} advantage across chemical space has yet to be found. While quantum computers may still prove useful for quantum chemistry, it may be prudent to assume exponential speedups are not generically available for this problem.
\end{abstract}

\section{Main}
The most common task in quantum chemistry is computing the ground electronic energy. The exponential quantum advantage hypothesis for this task is that for a large set of relevant (``generic'') chemical problems, this may be completed exponentially more quickly (as a function of system size) on a quantum versus classical computer\rev{~\cite{EQAQC}}. 
Here we examine this hypothesis.

We proceed using numerical experiments supported by theoretical analysis. 
To limit scope, we focus on fault-tolerant quantum algorithms --- the most advantageous setting for quantum computing, not limited by noise or hardware. 
Rigorously, computing the ground-state of even simple Hamiltonians can be exponentially hard on a quantum computer~\cite{kempe2006complexity}. 
However, such Hamiltonians might not be relevant to generic chemistry, and thus the specific exponential quantum advantage (EQA) hypothesis considered here, is that generic chemistry involves Hamiltonians which are polynomially easy for quantum algorithms (with respect to ground-state preparation) and simultaneously still exponentially hard classically, even using the best classical heuristics. 
Numerically, we thus focus on the evidence for quantum state preparation being exponentially easier than classical heuristic solution in typical problems; and whether the cost of classical heuristics in such problems scales exponentially with system size.
 
We stress that we do not attempt a rigorous proof or disproof of the EQA hypothesis, or other classes of quantum advantage.
Such proofs cannot be obtained, not least because what is ``generic'' chemistry is not precisely defined.
\rev{Our goal is to provide a coherent examination of} the evidence for whether exponential quantum speedup for ground-state determination in quantum chemistry should be considered likely in typical cases of interest ---  or, instead, very fine-tuned.

\section{Theoretical background}

\subsection{Statement of the problem} 

We compute the ground-state eigenvalue $E$ of the electronic Schr\"{o}dinger operator (Hamiltonian) of a chemical system discretized with a basis set, and the problem size is the basis size $L$. 
We consider the case where increasing $L$ corresponds to increasing physical system size (i.e.~number of atoms) with basis size  proportional to system size (other scenarios are discussed in the SI S1.1).  
The absolute ground-state energy $E$ increases with $L$, and in physical problems we expect extensivity (i.e.~$\lim_{L \to \infty} E(L) \propto L$ for a chemically uniform system);
in this limit, the energy density $\bar{E}=E/L$ may be the quantity of interest. Thus depending on the setting, the relevant error can be $\epsilon$ (error in $E$) or $\bar{\epsilon}$ (error in $\bar{E}$).

\subsection{Fault tolerant quantum algorithms for ground-state quantum chemistry} 

Fault tolerant quantum algorithms are ones which employ deep circuits (e.g.~depth is a function of $1/\epsilon$). The most famous one in quantum chemistry is quantum phase estimation (QPE)~\cite{kitaev1995quantum,aspuru2005simulated}. 
We focus on QPE for simplicity; qualitative features of the complexity remain similar in ``post-QPE'' algorithms~\cite{lin2021heisenberg}.
Phase estimation approximately measures the energy with approximate projection onto an eigenstate. The cost has 3 components (i)  preparing an initial state $\Phi$, (ii) the phase estimation circuit, and (iii) the number of repetitions (to produce the ground-state $\Psi_0$ rather than any eigenstate). 
The cost to obtain $E$ to precision $\epsilon$ is 
\begin{equation}
    \poly(1/S) [\poly(L) \poly(1/\epsilon) + C]
\end{equation}
where $C$ corresponds to (i), $\poly(L) \poly(1/\epsilon)$ corresponds to (ii), and $\poly(1/S)$ ($=1/S^{2}$ for QPE) with $S = |\langle \Phi | \Psi_0\rangle|$  corresponds to (iii). We term $S$ overlap and $S^2$ weight.

Motivated by the $\poly(L)$ cost of (ii), and assuming an $\exp(L)$ cost for classical solution, it is often informally stated that QPE yields EQA  
for the ground-state quantum chemistry task\rev{~\cite{EQAQC}}. 
However, the number of repetitions ($\poly(1/S)$) may also depend on $L$: this stems from the quality of state preparation. \rev{The restriction to generic chemistry effectively means  we assume that good state preparation is not exponentially hard due to unspecified additional structure. But such additional structure could also aid classical heuristics, and for EQA, the state preparation cost} must be exponentially less than the classical solution cost.

\subsection{State preparation}

\subsubsection{Ansatz state preparation}
We can prepare a state specified by an approximate classical ansatz. (We assume once an ansatz solution is specified, it is easy to prepare on the quantum device). 
\rev{Often,} 
simple states, such as the Hartree-Fock or Kohn-Sham ground-state (single Slater determinants) \rev{are considered in ansatz state preparation, as they are hoped to}
have good overlap with $\Psi_0$\rev{~\cite{o2019quantum,o2016scalable}; the $\poly(1/S)$ cost is then not further quantitatively considered. But while good overlap with such simple states can be} observed in small molecules, EQA is an asymptotic statement, thus we should consider the limit of large $L$. 

\rev{The orthogonality catastrophe~\cite{kohn1999nobel,chan2012low} has previously been discussed in the context of state preparation in the large $L$ limit~\cite{mcclean2014exploiting}. For a set of $O(L)$ non-interacting subsystems, the global ground-state is the product of the subsystem ground-states, thus if the local overlap between the approximate classical ansatz and ground-state for each subsystem is $\sim s < 1$, then the global overlap is $s^{O(L)}$ i.e. it decreases exponentially. This is sometimes viewed as an obstacle to ansatz state preparation, but in fact it does not rule it out; the issue is more subtle,
because the above analysis assumes that both the ansatz and the actual ground-state have some product structure. But one need not consider a classical ansatz with (approximate) product structure; and, at least in principle, ground-state correlations could mean that the global overlap is not guaranteed to be well approximated by a product of local overlaps. Also, even if one uses a product-like ansatz to approximate a ground-state of near-product form, one can improve the local overlap as a function of $L$, such that the global overlap is $1/\poly(L)$ or better.}

\rev{
The relevant consideration for EQA however, is that if classical heuristics can efficiently prepare states with such good overlap for large $L$ (for some systems), they may also efficiently obtain the ground-state energy to the desired precision. 
}



\subsubsection{Adiabatic state preparation (ASP)}
Alternatively, we can evolve slowly from the ground-state of a solvable initial Hamiltonian to that of the desired Hamiltonian\rev{~\cite{farhi2000quantum,albash2018adiabatic,veis2014adiabatic,aspuru2005simulated}}. This requires that the ground-state gap along the path be not too small;  for paths where the smallest minimum gap $\Delta_\text{min} \geq 1/\poly(L)$ (which we will term ``protected''), ASP plus QPE
provides a polynomial cost quantum algorithm. 
Since a protected gap is not guaranteed using an arbitrary initial Hamiltonian and path, ASP is a heuristic quantum algorithm. 
\rev{An extreme} problem that expresses the difficulty of finding a good path is unstructured search, where $\Delta_\text{min}$ acquires
a strong dependence on the ground-state $\Upsilon_0$ of the initial Hamiltonian, $\Delta_\text{min} \sim |\langle \Upsilon_0|\Psi_0\rangle|$~\cite{roland2003adiabatic}, yielding exponential cost when using adiabatic algorithms.

\rev{
The above raises several issues. First, in correlated quantum systems with competing ground-states, different phases could be separated by first-order phase transitions (where the gap is not protected) requiring ASP to start in the correct phase. Assuming one uses classical heuristics to prepare such a starting point and choice of path, one encounters similar questions to those raised in the discussion of ansatz state preparation. Second, one might ask how common the above situation is in generic chemistry, i.e. whether interesting chemical problems allow for initial Hamiltonians and paths with a protected gap to be trivially found.}

\subsection{The power of classical heuristics}

``Exact'' classical methods for ground state determination, such as exact diagonalization, are exponentially expensive on a classical computer (see SI S1.3). 
Thus the typical methods used in quantum chemistry are
classical heuristics, which come in a wide variety for different problems (see SI S1.2). 
The critical questions for EQA are (i) do these heuristics require $\exp(L)$ cost for given $\epsilon$ or $\bar{\epsilon}$ in their application domain,
(ii) does the patchwork of heuristics cover chemical space, and (iii) if there are gaps in coverage in practice, do we require classical methods of  $\exp(L)$ cost to cover them?

EQA assumes exponential-scaling cost of classical heuristic algorithms for given $\epsilon$ (or $\bar{\epsilon}$) across generic problems. We will examine this assumption in our numerical experiments. However, as actually employed in calculations, classical heuristics are often executed with $\poly(L)$ cost without necessarily achieving a specific accuracy, complicating the comparison with rigorous quantum algorithms. \rev{In particular, the error dependence can impact the EQA comparison, for example, a $\poly(L)\exp(\bar{\epsilon}^{-1})$ classical algorithm implies $\exp(L)$ cost for given $\epsilon$. Thus we will also examine the empirical precision dependence of classical heuristics with respect to $\epsilon$ or $\bar{\epsilon}$.}

\section{Numerical experiments}

\subsection{Fe-S clusters and metalloclusters of nitrogenase}

Iron-sulfur (Fe-S) clusters are amongst the most common bioinorganic motifs in Nature~\cite{beinert1997iron}, and the Fe-S clusters of nitrogenase have become a poster child problem for quantum chemistry on quantum devices~\cite{reiher2017elucidating,li2019electronic}. 
In the current context, they provide a concrete setting to assess the EQA hypothesis, in particular, the behaviour of quantum state preparation strategies. 

Specifically, we consider iron-sulfur clusters containing 2, 4, 8 transition metal atoms (the latter includes the famous FeMo-cofactor) in Figure 1. 
The 2, 4 metal clusters will be referred to as [2Fe-2S], [4Fe-4S] clusters, while the 8 metal clusters include the P-cluster (8Fe) and the FeMo-cofactor (7Fe, 1Mo).
\rev{We note that the P-cluster and FeMo-cofactor are the largest Fe-S clusters found in Nature.}
We represent the electronic structure in active spaces with Fe 3d/S 3p character constructed from Kohn-Sham orbitals. Within the occupation number to qubit mapping, this corresponds to up to $40$ qubits ([2Fe-2S]), up to $72$ qubits ([4Fe-4S]), and up to $154$ qubits (P-cluster and FeMo-co) (see SI S3.1). 
For [2Fe-2S], exact solutions can be obtained using exact full configuration interaction (FCI).
For all clusters, we obtain a range of 
approximate solutions using the quantum chemistry density matrix renormalization group (DMRG) \rev{\cite{white1999ab, chan2011density, baiardi2020density, sharma2014low, li2019electronic2}} with a matrix product state (MPS) bond dimension $D$; increasing $D$ improves the approximation, allowing extrapolation to the exact result (see SI S3.3).  
Note that the classical calculations in this section are of interest mainly to provide data to understand quantum state preparation.

\subsubsection{Nature of the ground-state and cost of ansatz state preparation} 

We first examine the nature of the ground-state $\Psi_0$ and the cost of ansatz state preparation. For this, we compute the weight of a Slater determinant $S^2 = |\langle \Phi_D|\Psi_0 \rangle |^2$, shown in Figure 1{\fontfamily{phv}\selectfont \textbf{B}}. 
$\Phi_D$ is parametrized by its orbitals $\{ \phi \}$, and we choose a priori, or optimize, $\{ \phi\}$ to maximize this weight (for a best-case scenario that uses information from the solution $\Psi_0$, see SI S3.4).  
The weights decrease exponentially  over a small number of metal centers, and are already very small in FeMo-co ($\sim 10^{-7}$). 
The number of QPE repetitions is $\poly(1/S)$, yielding a large prefactor even for an ``optimized'' Slater determinant. 

\begin{figure*}[!htp]
  \centering
  \includegraphics[width=.86\linewidth]{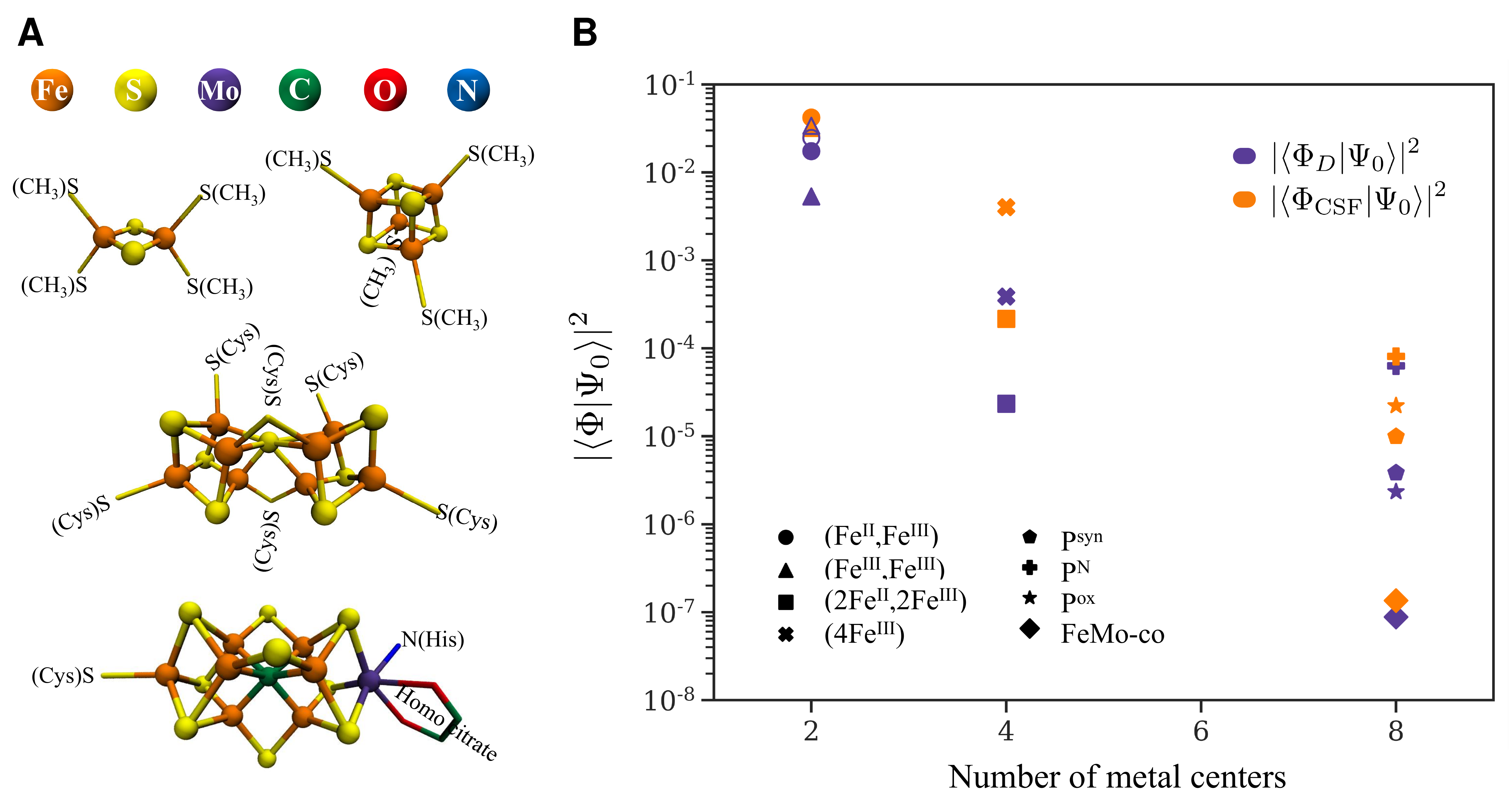}
\caption{ \textbf{Ansatz state preparation and ansatz weights for model Fe-S clusters.} {\fontfamily{phv}\selectfont \textbf{A,}} Structural models of [2Fe-2S], [4Fe-4S], P-cluster, and FeMo-co. {\fontfamily{phv}\selectfont \textbf{B,}}
Weight of two different types of ansatz state: largest weight determinant ($\Phi_D$) (purple) 
and largest weight configuration state function ($\Phi_\mathrm{CSF}$) (orange)
as a function of the number of metal centers in each cluster (using split-localized orbitals). P$^\text{N}$, P$^\text{syn}$, P$^\text{ox}$ here refer to different oxidation states of the metal ions in the P-cluster.
Both types of ansatz state show an exponential decrease in weight with the number of metal centers. For the [2Fe-2S] clusters, we also show results for the largest weight determinant using natural orbitals (empty symbols).}
\label{fig:structure_overlap}
\end{figure*}

We next prepare more complex states with better overlap. We use a single configuration state function (a linear combination of Slater determinants that is an eigenfunction of total spin\rev{~\cite{helgaker2014molecular}}). The weights improve but still show exponential decay \rev{to small values.}

\rev{
These results indicate that the magnitude of the ansatz overlap can become
a relevant concern even in systems of modest size when using ansatz state preparation, thus we should consider improved state preparation. Assuming the ground-state is of approximately product form, we can obtain some rough insight into improved global ansatz state preparation (e.g. for 8 metal clusters) from the behaviour of improving the state locally (i.e. for the 2 metal clusters); we require $\poly(1/I)$ cost for the [2Fe-2S] fragment ($I$ is the infidelity $1-S^2$) for efficient global state preparation.  In SI S2.1 
we show that this cost is indeed $\poly(1/I)$, but also that the energy error is $\poly(I)$. Thus under these assumptions, improving the local overlap sufficiently also implies efficient classical solution for the global energy. For any finite system, it may be possible to engineer a practical quantum advantage for some target precision from a sufficiently good ansatz overlap and a favourable ratio of classical and quantum costs. 
But the problem of finding a classical heuristic that efficiently yields $1/\poly(L)$ overlap but which cannot also efficiently reach the target precision remains.
}



\subsubsection{Adiabatic state preparation}

We next compute the ASP cost for a simplified $n_\text{act}=12$ active space (24 qubit) [2Fe-2S] model  (see SI S4.1). 
The path is a heuristic input, 
and we use one which linearly interpolates the Hamiltonian $H(s)$ (with ground-state $\Upsilon_0(s)$) between an initial Hamiltonian ($s=0$, with ground-state $\Upsilon_0(0)$) and the true Hamiltonian ($s=1$, with ground-state $\Upsilon_0(1)\equiv \Psi_0$); the path preserves spin symmetry. We consider two families of $H(0)$; a set of mean-field Hamiltonians (with different Slater determinant ground-states) and a set of interacting Hamiltonians (these retain interactions among $q$ active spin-orbitals (qubits), definitions in SI S4.2). 

Tight bounds on the ASP time ($T_\text{ASP}$) are difficult to obtain (see SI S1.5). 
However, we have verified that the adiabatic estimate $T^\text{est}_\text{ASP} \sim \max_s \tau(s)$, with $\tau(s) = |\langle \Upsilon_0(s)|dH(s)/ds| \Upsilon_1(s)\rangle|/\Delta^2(s)$
with $\Delta(s)$ the ground-state gap and     $\Upsilon_1(s)$ the first excited state of $H(s)$, is a reasonable estimate for the desired preparation fidelity (here assumed 75\% final weight) by carrying out time-dependent simulations of ASP for simple instances to compute $T_\text{ASP}/T^\text{est}_\text{ASP}$ (Figure 2{\fontfamily{phv}\selectfont \textbf{B}}); 
for a range of examples, this ratio is $O(1)$. Thus we use $T^\text{est}_\text{ASP}$ as the ASP time below.

Figure 2{\fontfamily{phv}\selectfont \textbf{C}} 
shows $T_\text{ASP}$ across the sets of $H(0)$;  it varies over 8 orders of magnitude depending on the choice of $H(0)$. 
We see a trend $1/(\min_s \Delta(s)) \sim \poly(1/|\langle \Upsilon_0 | \Psi_0 \rangle|)$ and thus
$T_\text{ASP} \sim \poly(1/|\langle \Upsilon_0 | \Psi_0 \rangle| )$ 
reminiscent of unstructured search. The practical consequence is that an a priori good choice of initial Hamiltonian is non-trivial; the mean-field Hamiltonian with the lowest ground state energy gives a large $T_\text{ASP} > T_\text{QPE}$ (\rev{an estimate of the total} coherent QPE evolution time for 90\% confidence, $\epsilon$=$10^{-3}E_h$, \rev{see analysis in SI S2.5), } 
while out of the interacting $H(0)$'s, we need to include almost all the interactions when diagonalizing $H(0)$ for the initial state (20 out of 24 qubits) before $T_\text{ASP}<T_\text{QPE}$.  
Although these results are for the smallest FeS cluster, the dependence of $T_\text{ASP}$ on $S$ is problematic for EQA should it scale to larger interesting problems,
\rev{and it illustrates the importance of heuristics to find a 
good initial starting point for ASP in relevant chemical problems. As discussed above, if classical heuristics are used for this task, this raises the question of whether they are exponentially advantageous over the classical heuristics for solution.}


\begin{figure*}[ht!]
\centering
\includegraphics[width=0.8\textwidth]{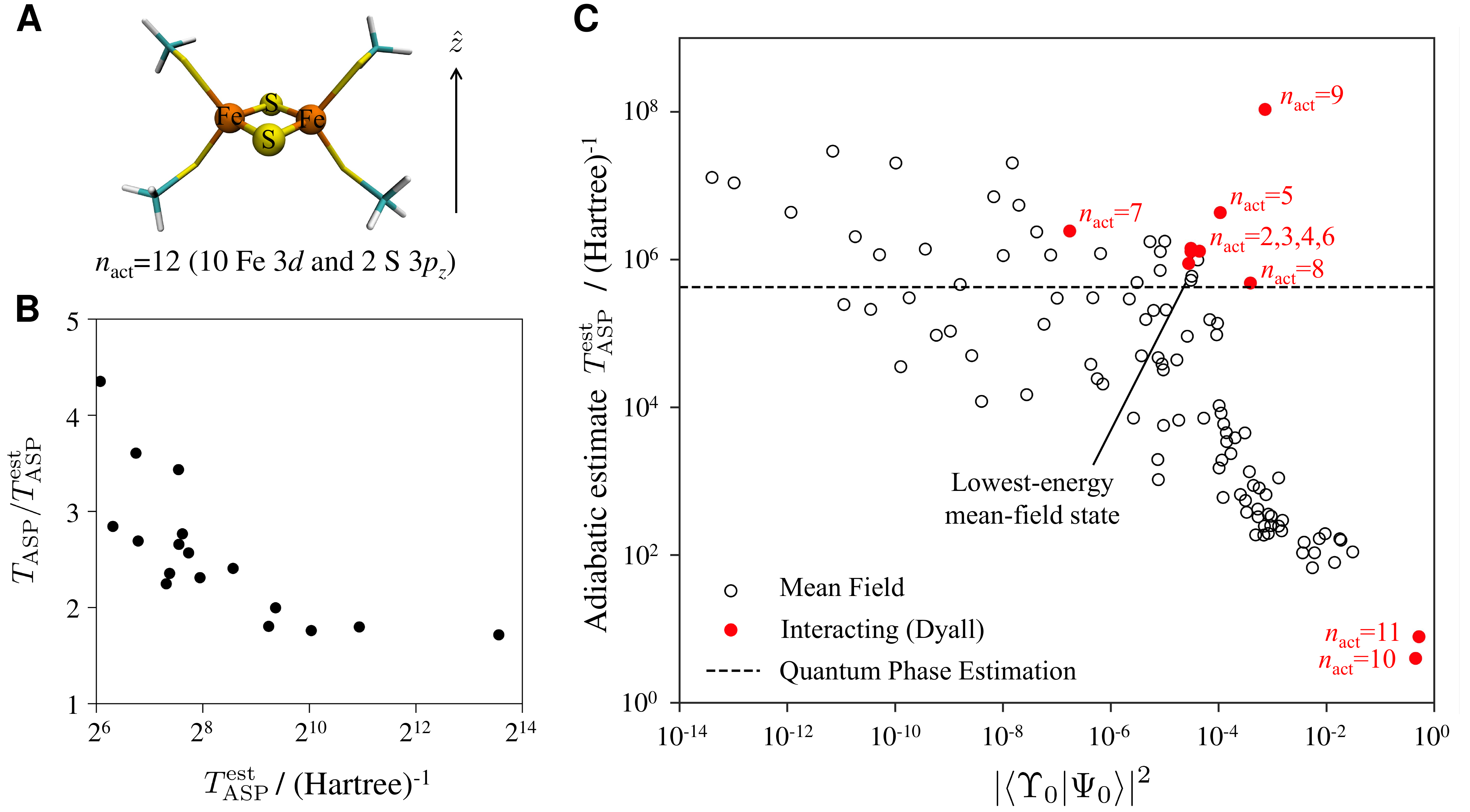} 
\caption{ \textbf{Adiabatic state preparation for a model [2Fe-2S] cluster.} {\fontfamily{phv}\selectfont \textbf{A,}} Structure and simplified active space model of [2Fe-2S] cluster. 
{\fontfamily{phv}\selectfont \textbf{B,}} ASP time and the adiabatic estimate. We see that the ratio $T_{\mathrm{ASP}}/T^\mathrm{est}_\mathrm{ASP}$ is $O(1)$. 
{\fontfamily{phv}\selectfont \textbf{C,}} Adiabatic estimates ($T^\mathrm{est}_\mathrm{ASP}$) for two families of initial Hamiltonians against the weight of the initial ground state ($\Upsilon_0$) in the final ground state ($\Psi_0$) ($|\langle \Upsilon_0|\Psi_0\rangle|^2)$, showing an inverse dependence on the initial weight. The mean-field Hamiltonians are constructed to have different Slater determinants as their ground-state, while the interacting Hamiltonians contain the full electron interaction amongst $n_\text{act}$ orbitals.
Additional discussion in SI S4 and S2.3.
}\label{fig:asp}
\end{figure*}

\subsection{The scaling of classical heuristics}

The Fe-S cluster simulations raise questions as to whether high quality quantum state preparation \rev{can be assumed to be} exponentially easier than classical heuristic solution. We now consider if classical heuristics in fact display 
$\exp(L)$ cost for fixed $\epsilon$ or $\bar{\epsilon}$, as is required to establish EQA.  We do so by considering examples that arguably represent much of chemical space, which are evidence of classical heuristics scaling to large problems and high accuracy at polynomial cost for fixed $\bar{\epsilon}$. (If the error scaling is $\poly(1/\bar{\epsilon})$ independent of $L$, this further implies $\poly(L)$ cost overhead to achieve fixed $\epsilon$). We note that the calculations below represent only a small slice of relevant evidence from classical calculations; \rev{related calculations can be found in the literature, although our focus here is on characterizing the calculations e.g. with respect to cost and precision in a way  useful for understanding EQA.} 
Some further discussion of these systems and other calculations relevant to EQA is in SI S2.6. 

For ``single-reference'' chemical problems (see SI S5) 
coupled cluster (CC) wavefunctions  
are often described as the gold-standard. The heuristic assumes that correlations of many excitations can be factorized into clusters of fewer excitations; if the maximal cluster excitation level is independent of $L$, the cost is $\poly(L)$   (assuming a non-exponential number of iterations for solution) without guaranteed error. To establish the error dependence, Figure 3{\fontfamily{phv}\selectfont \textbf{A}} 
shows the empirical convergence of error as a function of cost, consistent with $\poly(1/\epsilon)$ scaling, for a small molecule (\ce{N_2}).  By the extensivity of the coupled cluster wavefunction, this translates to $\poly(L)\poly(1/\bar{\epsilon})$ cost for a gas of non-interacting $\ce{N_2}$ molecules, and thus $\poly(L)\poly(1/\epsilon)$ given the error convergence above. 
We can take this as a conjectured complexity of coupled cluster in single-reference problems. To practically test this scaling form on larger systems, we introduce another heuristic. 
CC methods can be formulated to exploit locality, a commonly observed and widely conjectured feature of physical ground-states (including gapless systems, see SI S1.4). 
This yields the local CC heuristic that has cost linear in $L$ in gapped systems~\cite{VRG:riplinger:2016:JCP,yang2014ab}. 
Figure 3{\fontfamily{phv}\selectfont \textbf{B}} 
illustrates the application of local CC to $n$-alkanes, a set of organic molecules, with the associated computational timing. This suggests the cost is $O(L)$, while the computed  enthalpy of formation per carbon achieves constant error versus~experimental data, reflecting constant $\bar{\epsilon}$ as a function of $L$, consistent with the conjectured complexity. 
Many biomolecules are single-reference problems, allowing local coupled cluster methods to be applied to protein-fragment-scale simulations (Figure 3{\fontfamily{phv}\selectfont \textbf{C}}). 

\begin{figure*}[!htp]
  \centering
  \includegraphics[width=.85\linewidth]{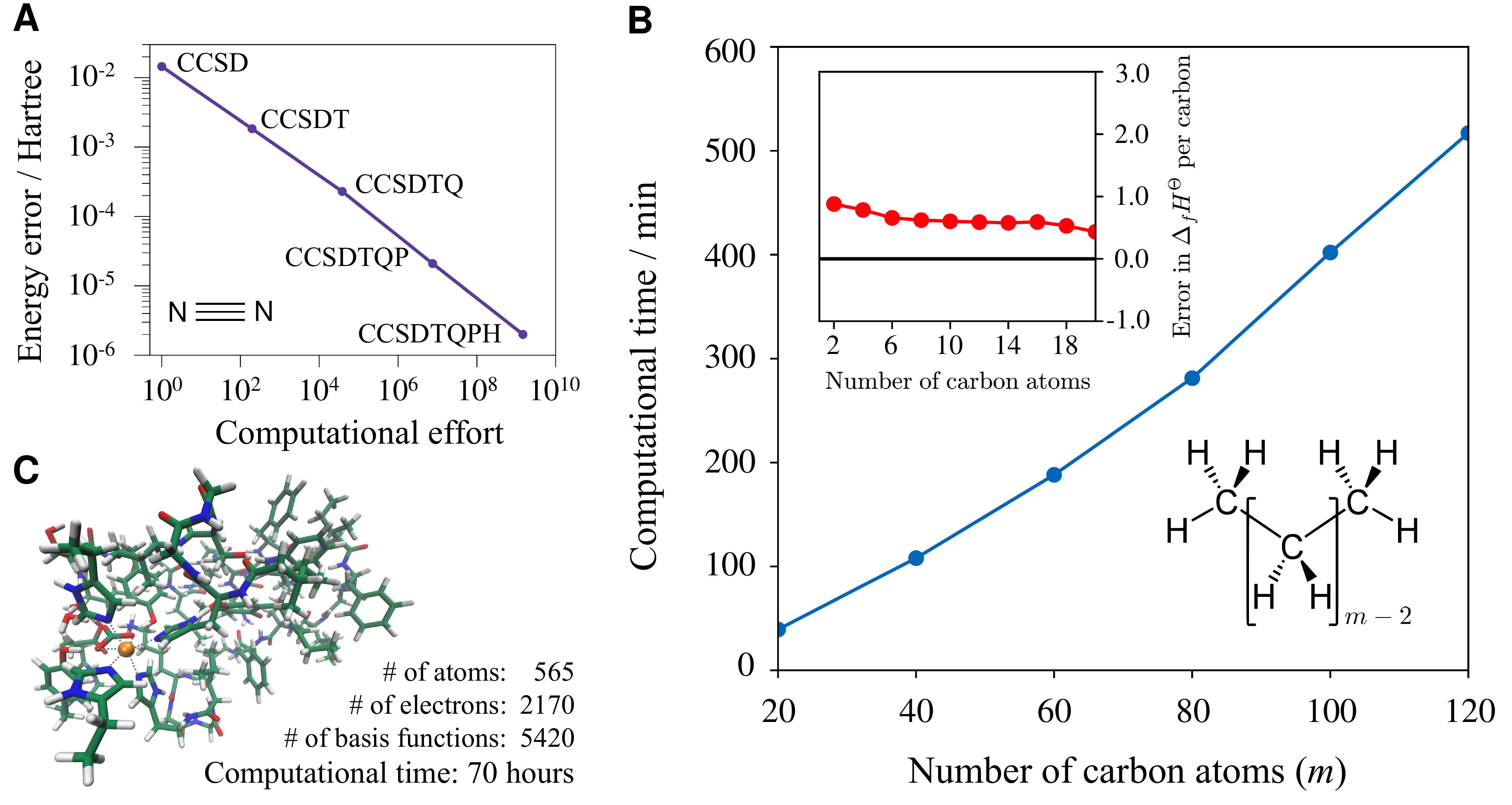}
\caption{ \textbf{Computational complexity of classical heuristics for molecular systems.} {\fontfamily{phv}\selectfont \textbf{A,}} Energy error of a nitrogen molecule (equilibrium geometry) as a function of the level of CC approximation, against a computational time metric. Data taken from Ref.~\cite{chan2004state}, time metric defined in SI S2.7. 
The observed precision cost is like $\poly(1/\epsilon)$.
{\fontfamily{phv}\selectfont \textbf{B,}} Cost of a state-of-the-art reduced-scaling coupled cluster (CCSD(T)) implementation scales nearly-linearly with the system size in gapped systems, as demonstrated here for $n$-alkanes (C$_m$H$_{2m+2}$) with with $m=[20 \ldots 120]$. Size-extensivity of the coupled-cluster ansatz ensures constant error per system subunit, as illustrated in the subfigure for the error of explicitly-correlated reduced-scaling CCSD(T)\cite{VRG:kumar:2020:JCP} (see SI S5.1 
for details) with respect to the available experimental gas-phase enthalpy of formation in the standard state for $n$-alkanes with $m=[2 \ldots 20]$.
{\fontfamily{phv}\selectfont \textbf{C,}} Reduced-scaling CCSD(T) implementations can be routinely applied to systems with thousands of electrons on a few computer cores, as demonstrated here for a small fragment of photosystem II. \cite{VRG:muh:2013:PR}} \label{fig:lccsd}
\end{figure*}

\begin{figure*}[!htp]
  \centering
  \includegraphics[width=.85\linewidth]{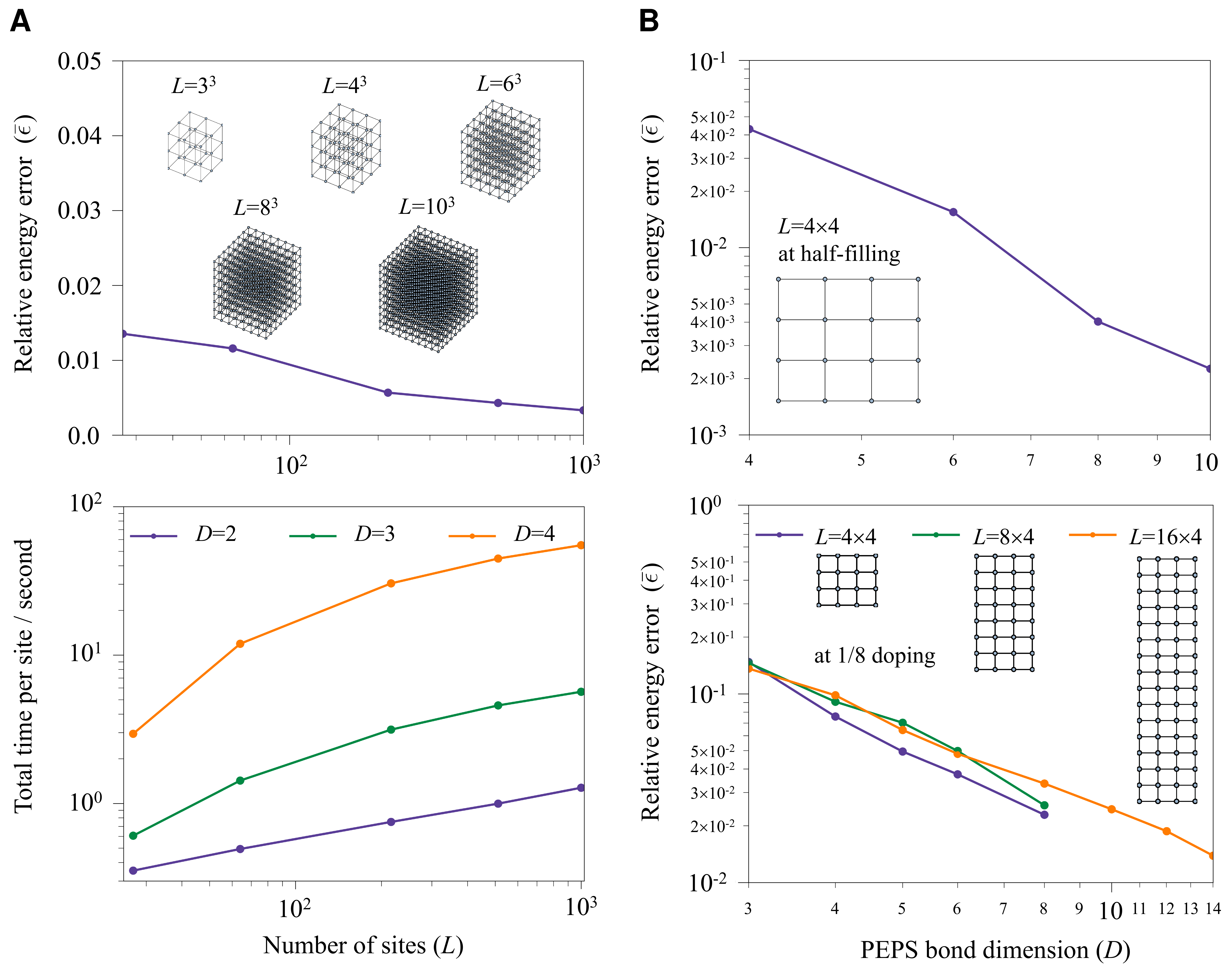} 
\caption{ \textbf{Computational complexity of classical heuristics for models of strongly correlated material systems.} {\fontfamily{phv}\selectfont \textbf{A,}} Relative energy error of a tensor network (PEPS) with respect to system sizes $3^3$ to $10^3$ for the 3D Heisenberg cube model with a bond dimension ($D$) of 4. In bottom panel: total computational time in seconds, divided by number of sites, as a function of system size, demonstrating $\poly(L)$ (close to linear) computational effort.
{\fontfamily{phv}\selectfont \textbf{B,}} Energy convergence of PEPS with respect to the bond dimension for  2D Hubbard models at half filling ($4\times4$ lattice, in top panel) and the challenging 1/8 doping point ($4\times4$, $8\times4$ and $16\times4$ lattice, in bottom panel). The plots are consistent with $1/\bar{\epsilon}\sim \poly(D)$ with a weak dependence on $L$.} \label{fig:tns}
\end{figure*}

Strongly correlated materials (e.g.~bulk analogs of the Fe-S clusters) remain challenging to treat with ab initio quantum chemistry (although there has been considerable progress in recent years~\cite{motta2017towards,cui2021systematic}). 
To obtain insight into the computational complexity, 
it is more practical to study simpler models of correlated materials (e.g.~the Heisenberg and Hubbard models, often used to study quantum magnets and  high temperature superconductors~\cite{arovas2021hubbard,zheng2017stripe}). 
Many methods can now access large parts of these model phase diagrams to reasonable accuracy without $\exp(L)$ cost. 
The use of locality is common to several heuristics for strongly correlated problems; tensor networks are an example of such a class of heuristics and we examine illustrative applications below. (Note that this is not an exhaustive study of tensor networks, nor of other heuristics (such as quantum embedding); for additional discussion see SI S2.6). 

Figure 4 
shows results from a tensor network ansatz~\cite{orus2019tensor}, where the expressiveness of the ansatz is controlled by the bond dimension $D$. The contraction here is explicitly performed with  $\poly(D)$ (typically a high polynomial) cost, thus for given $D$ (assuming the number of iterations is not exponential in $D$ or $L$ (see SI S6)), 
the algorithm cost is $\poly(L)\poly(D)$,  without guaranteed error.
$\bar{\epsilon}$ and computational cost are shown as a function of $L$ in the 3D cubic Heisenberg model, and $\bar{\epsilon}$ as a function of $D$ and $L$ in the 2D Hubbard model. 
(Note: these examples were chosen for ease of generating exact data, rather than representing the limits of classical methods in size, accuracy, or complexity of physics; see SI S2.6 
for other examples).  Figure 4{\fontfamily{phv}\selectfont \textbf{A}} 
shows that the cost is close to $O(L)$ in the 3D Heisenberg model for up to 1000 sites, while achieving close to constant $\bar{\epsilon}$.
Less data is available for the error scaling as accessible $D$ remain small; in particular it is currently too expensive to reach large enough $D$ to meaningfully study the $\bar{\epsilon}$ scaling in 3D. However in the 2D Hubbard model (Figure 4{\fontfamily{phv}\selectfont \textbf{B}}) 
we see $\bar{\epsilon} \sim 1/\poly(D)$ (or slightly better)
across a range of studied $D$, (with a weak dependence on $L$) even at the challenging 1/8 doped point of the model. Assuming this error form holds asymptotically, the observed empirical cost is  $\poly(L)\poly(D)\poly(1/\bar{\epsilon})$, which corresponds to $\poly(L)\poly(D)\poly(1/\epsilon)$ for the assumed error scaling, and we can conjecture that this holds also in 3D. 

\rev{Although the Hubbard and Heisenberg models are believed to contain the basic physics of many strongly correlated materials, moving from such simplified models to more detailed quantum chemistry models (i.e. \textit{ab initio} Hamiltonians) will certainly increase complexity. But establishing EQA requires evidence that adding the polynomial number of terms in the Hamiltonian causes the classical heuristic to fail or become exponentially expensive. 
The history of development of classical heuristics does not support this, as methods originally developed on simpler models routinely graduate to \textit{ab initio} simulations. For example, the coupled cluster methods described above were first developed for use in model simulations, as were simpler tensor networks such as the density matrix renormalization group and tree tensor networks now used in \textit{ab initio} calculations~\cite{white1999ab,nakatani2013efficient,mayhall2017using}; SI S2.6 
provides more discussion of this point as well as shows the performance of a quantum embedding heuristic for the 2D and 3D hydrogen lattices, \textit{ab initio} analogs of the 2D Hubbard and 3D Heisenberg systems in Figure 4.} 
Further examples in the literature consider the application of many different classical heuristics to ab initio or model chemical ground-states of complex systems including strongly correlated materials~\cite{motta2017towards,cui2021systematic,li2019electronic2,williams2020direct,yang2014ab,VRG:kumar:2020:JCP,brandenburg2019physisorption,zheng2017stripe,qin2020absence}. Although the computational complexity is not formally analyzed, the success of such studies of large and complex problems is compatible with the view
that the ground-state quantum chemistry problem is often soluble with classical heuristics, to an energy density error $\bar{\epsilon}$ relevant to physical problems, with $\poly(L)$ cost. 
Thus, while there are many  chemistry problems that cannot currently be addressed by classical methods {and further study can be expected},
the barrier to solution may be of polynomial (even if impractically large) rather than exponential cost.

\section{Conclusions}
We have examined the case for the exponential quantum advantage (EQA)  hypothesis for the central task of ground-state determination in quantum chemistry. The specific version of EQA that we examined required quantum state preparation to be exponentially easy compared to classical heuristics, and for classical heuristics to be exponentially hard. Our numerical simulations
\rev{highlight that heuristics are necessary to achieve efficient quantum ground-state preparation. At the same time, we do not find evidence for the exponential scaling of classical heuristics in a set of relevant problems. The latter suggests that quantum state preparation can be made efficient for the same problems. However, as EQA is based on the ratio of costs, this does not lead to EQA.
}


Numerical calculations are not mathematical proof of asymptotics with respect to size and error, nor can we exclude EQA in specific problems. However, our results suggest that 
without new and fundamental insights, there may be a lack of generic EQA in this task. \rev{Identifying a relevant quantum chemical system  with strong evidence of EQA remains an open question. }

We did not consider tasks other than ground-state determination, nor do we rule out polynomial speedups. \rev{Depending on the precise form,} polynomial \rev{quantum} speedups could be associated with useful quantum advantage, \rev{as even a polynomial classical algorithm does not mean that solutions can be obtained in a practical time}. Both aspects may prove important in the further development of quantum algorithms in quantum chemistry. \rev{For further discussion, we refer to the FAQ (see Sec.~\ref{sec:FAQ}).}

\section{Acknowledgement}
Work by SL, HZ, GKC was funded by the US Department of Energy, Office of Science, via Award DE-SC0019374. Work by PH was funded by the Simons Collaboration on the Many-Electron Problem, and support from the Simons Investigator Award to GKC. Work by ZC was funded by the US Department of Energy, Office of Science, via Award DE-SC0019390. Work by JP was funded by the US Department of Energy, Office of Science, via Awards DE-NA0003525, DE-SC0020290, and by the National Science Foundation via Award PHY-1733907. Work by YT was funded by the US Department of Energy, Office of Science via Award DE-SC0017867.
Work by LL was funded by the National Science Foundation via Award OMA-2016245, and by the Simons Investigator Award.
Research by AK and EV was funded by the US Department of Energy, Office of Science, via Award DE-SC0019374, and the associated software development efforts were supported by the US National Science Foundation via Award OAC-1550456. \rev{Some of the discussions and collaboration for this project occurred while using facilities at the Kavli Institute for Theoretical Physics, supported in part by the National Science Foundation under Grant No. NSF PHY-1748958.}
RB thanks members of the Google Quantum AI team for helpful feedback on earlier drafts.

\section{Frequently asked questions (FAQ)} \label{sec:FAQ}
\begin{enumerate} 

\item \textit{Are there any published papers which suggest exponential quantum advantage for ground-state quantum chemistry?} Statements of this kind can be found in various settings, and range from direct statements of EQA for ground-state problems, to more implicit statements where the expectation of EQA could reasonably be inferred by the reader.  Some representative papers and further discussion can be found in Ref.~\cite{EQAQC}.

\item \textit{Does this mean that quantum computers are not useful for generic ground-state quantum chemistry?} Even if EQA is not found in the ground-state problem, quantum computers may still be useful for this task, since polynomial advantage (or even large constant factor advantage) can be very useful. In such cases, the details (e.g. degree of polynomial, size of constant) for both the classical heuristic and quantum algorithm are important. It is difficult to tie generic statements about advantage to other characteristics, such as strength of correlation; for example, in more strongly correlated systems, state preparation must be more carefully considered due to possible competing phases.
However, in terms of assessing the suitability of ground-state quantum chemistry as an early target application for quantum computers, one should compare the degree of available polynomial advantage in quantum chemistry to that in other applications.  

\item \textit{We do not have answers to a certain problem by classical heuristics, and exact classical solution is exponentially expensive, doesn't this mean that there should be EQA?} As our work indicates, this is a subtle question. 
First, one can definitely construct artificial ground-state problems which, for certain precision requirements, can be solved efficiently quantumly but not classically. See e.g. SI S1.3 
as well as Refs.~\cite{gharibian2021dequantizing,cade2022complexity}. The open question there is whether such problems are related to the generic molecules and materials studied in quantum chemistry. Second, there are relevant chemical problems which are too large to treat with current classical heuristics with the desired precision. However, for EQA, one must establish that e.g. to reach the desired precision, the classical heuristic requires exponential effort as a function of system size, 
and also that quantum algorithms are capable of reaching the desired precision without exponential effort. 

\item \textit{What about potential improvements to quantum algorithms?} We cannot anticipate future improvements to quantum algorithms, for example, in the area of quantum heuristics. However, to change the situation regarding EQA it is critical for an improvement to change the ratio of the quantum to classical costs. For example, when introducing locality into a quantum heuristic, we should understand whether classical heuristics that also use locality are efficient for the same problems.

\item \textit{What about other classes of problems such as quantum dynamics for chemical systems?}
We cannot conclude anything  about other simulation tasks based on this work.
However, we note that other tasks, such as the simulation of chemical dynamics, may also be amenable to heuristics because of the particular chemical setting and question of interest; for example, dynamics of heavier atoms can often be treated classically, or certain phenomena may take place with strong dissipation. Thus heuristics should always be carefully considered in claims of EQA and other types of quantum advantage. 

\end{enumerate} 

\bibliographystyle{naturemag}
\bibliography{ref_SI,ref}



\section*{Supplementary material}

# Contents



# S1 Theoretical background

## S1.1 Basis size scaling

In the main text, we described scaling with $L$ where $L$ is the physical system size. However, we can also consider increasing the basis size $L$ while keeping the physical system size fixed. One can view this as increasing the spatial resolution with which one is representing the system. This type of scaling exhibits very different characteristics to scaling the physical system size. For example, we expect the energy to (smoothly) converge to the basis set limit result, a finite number (i.e. $\lim_{L\to\infty} E(L) \to E(\infty)$). For a fixed error $\epsilon$ (with respect to $E(\infty)$), the computational cost should become independent of $L$ for sufficiently large $L$, as one can simply neglect any additional basis functions. Similarly, past a certain basis size, we do not expect the ground-state to change much, thus the state preparation cost can also become independent of $L$ for sufficiently large $L$. For the chemical Hamiltonian, the asymptotic rate of convergence of the exact $E(L)$ is known to be controlled by the singularity in the electron-electron interaction at zero-separation which creates a cusp in the wavefunction along the electron-electron coordinate, see e.g. Ref. [1].

## S1.2 Overview of classical heuristics

There are many types of classical heuristics for quantum chemistry and quantum many-body problems. They are too numerous to describe here, we only give a short overview with some references. A non-exhaustive list includes empirically parameterized methods, such as density functional methods [2] (which are generally of modest cost but which are not systematically improvable) to non-parametrized many-body methods, which usually have some computational parameter which may be tuned such that $\epsilon \to 0$. Out of the latter, there are methods based on perturbing around mean-field wavefunctions (this includes standard coupled cluster theory) [3], various flavours of quantum Monte Carlo methods [4, 5], systematically improvable variational

wavefunctions, such as configuration interaction wavefunctions [3], tensor network states [6, 7], and neural network wavefunctions [8, 9], and reduced density matrix approaches based on $N$-representability constraints [10]. Each method makes tradeoffs to avoid the exponential scaling of exact methods, and the quality of these tradeoffs defines different optimal domains of application. Recent years have seen benchmark studies comparing a wide variety of different heuristic methods, see e.g. Refs. [11, 12, 13].

## S1.3 Complexity of rigorous quantum and classical methods for quantum chemistry under the assumption of efficient state preparation

The main text highlights the importance of both quantum and classical heuristics in establishing EQA, and that quantum advantage in practice is related to the relative cost of (heuristic) quantum state preparation versus classical heuristics for the ground-state problem.

If, however, we separate the problem of state preparation from that of ground-state determination (i.e. we assume that a good initial state is available, which may or may not be efficiently preparable by classical or quantum means) then it is possible to establish conditions which demonstrate a clear separation between rigorous quantum and classical algorithms, i.e. algorithms with error guarantees.

For example, in estimating the ground state energy, QPE and other related quantum algorithms have a rigorous precision guarantee under the assumption of having an initial guess with good overlap. If we require classical methods to have the same rigorous precision guarantee, under the same assumption, then we can construct problems where quantum algorithms can have a very significant advantage. Here we consider the setting where we have a good initial guess $|\Phi_0\rangle$ for the ground state $|\Psi_0\rangle$ with overlap $S = |\langle \Phi_0 | \Psi_0 \rangle|$ that is lower bounded by $1/\operatorname{poly}(L)$, where $L$ is the system size.

On a classical computer, brute-force diagonalization to compute the ground-state energy will

result in $\exp(L)$ runtime and memory requirements. However, there are more clever ways to obtain the ground state energy. One can compute quantities of the form $\|f(H)|\Phi_0\rangle\|$ for polynomials $f$, and this enables us to classically implement the binary search procedure described in Ref. [14]. With this method, let us consider a setting where the Hamiltonian, relative to a suitable computationally accessible basis, has $\text{poly}(L)$ non-zero elements in each row and column, while $|\Phi_0\rangle$ contains $\mathcal{O}(1)$ non-zero elements relative to the same basis. The number of matrix-vector multiplications needed is $\text{polylog}(\|H\|\epsilon^{-1})$, corresponding to the sum of the degrees of the polynomials $f$ we need to implement. Each time matrix-vector multiplication is performed, the number of non-zero elements of the vector grows by a factor of $\text{poly}(L)$ which corresponds to the sparsity of the Hamiltonian. The total runtime will in the end be $\text{poly}(L)^{\text{polylog}(\|H\|\epsilon^{-1})}$, where $\|H\| = \text{poly}(L)$. In a similar setting, QPE and other related quantum algorithms will have a $\text{poly}(L)\epsilon^{-1}$ runtime. Therefore at least for certain Hamiltonians and guarantees about the initial state, there can be a significant advantage for a quantum algorithm versus a classical algorithm with the same rigorous error guarantee.

The appearance of $\text{polylog}(\|H\|\epsilon^{-1})$ in the above exponent suggests that, assuming the availability of an initial state with good overlap, there is hope of having a polynomial time classical algorithm with guaranteed performance for the ground state energy only when constant *relative* precision $\|H\|\epsilon^{-1}$ (which, if $\|H\| = \mathcal{O}(L)$ means constant precision in $\bar{\epsilon}$) is required. Indeed, such a result was recently obtained (Theorem 1 in Ref. [15]) which established that estimating the ground state energy of a local Hamiltonian to constant relative precision can be achieved in polynomial time on a classical computer, assuming sampling access to an initial guess of the ground state. Note that we did not require computing a description of the ground state to any precision. If the ground state is required, then using methods such as the Lanczos iteration, a relative spectral gap dependence (relative to $\|H\|$) will need to be introduced, making the runtime scale like $\text{poly}(L)^{\text{poly}(\|H\|)\text{polylog}(\|H\|\epsilon^{-1})}$. For this task, the runtime of QPE and other

related quantum algorithms will still be $\text{poly}(L)\epsilon^{-1}$.

Ref. [15] established another separation between rigorous quantum and classical algorithms. Assuming access to a good initial guess of the ground state, Theorem 2 of Ref. [15] showed that estimating the ground-state energy of certain Hamiltonians to inverse-polynomial absolute (and relative) precision is BQP-hard. The proof proceeds by using a circuit-to-Hamiltonian construction [16] to encode a polynomial-time BQP computation into a local Hamiltonian, where the ground state energy is related to whether the answer to the BQP problem is yes or no; because the relevant gap of the problem vanishes inversely polynomially with system size, the ability to estimate the energy to inverse polynomial precision using a classical algorithm would imply BQP = BPP, which is generally regarded as unlikely. Certain aspects of this construction may not necessarily be common in physical problems. For example, although there has been considerable effort to make the Hamiltonians involved more physically reasonable (for example, it has been shown that they can be composed of nearest-neighbor interactions on a 1D (high-local dimension) spin chain [17], and these interactions can even be translationally invariant [18]) it remains unclear whether they arise naturally in physical settings. In general, the relevant ground states are critical systems whose spectral gaps close inverse-polynomially, and moreover, by construction, their ground states are highly entangled (as they encode the states encountered by an arbitrary quantum computation). This contrasts with generic systems in quantum chemistry which might be expected to obey an area law (see SI S1.4). This aspect, along with the need for inverse-polynomial absolute precision on the energy estimate [19], complicates any concrete conclusions that can be drawn from Ref. [15] about the prospect of rigorous EQA in quantum chemistry, even under the assumption of efficient state preparation.

## S1.4 Locality and area laws

**Locality and entanglement area laws**. The extensive energy as $L \to \infty$ is a result of the long-range decay (locality) of interactions between electrons and the charge neutrality of the system. This type of locality might be termed Hamiltonian locality (and is closely related to the more formal idea of geometric locality in lattice systems).

A related statement of locality for quantum states is the area "law" of entanglement for ground-states (strictly a conjecture in the general setting) which states the entanglement entropy for a bipartition is proportional to the length of the boundary, with at most logarithmic violations for critical states (e.g. metallic states). Locality is believed to be a characteristic of low-energy physical eigenstates, but is not itself sufficient to ensure that the resulting quantum chemistry problem is easy for quantum or classical simulation. Intuitively, this is due to the possibility of glassiness or frustration, where additional local interactions as $L$ increases can drastically change the global ground-state.

**Gaps**. Another useful concept is the gap $\Delta$ between the ground-state and first excited state, and the related density of states near the ground-state. Many formal and empirical complexity results depend on the magnitude of this gap. For example, the area law in 1D can be proven for local Hamiltonians for constant $\Delta$ [20, 21], while the area law holds in all dimensions for constant $\Delta$ and a polynomial density of states near the ground-state [22].

**Provably efficient algorithms in 1D**. The power of area laws and bounded gaps is clearly seen in 1D, where there is a provably $\text{poly}(L)$-time classical tensor network algorithm for the ground state energy problem, assuming the Hamiltonian has a constant gap and all interactions are nearest neighbor on a 1D spin chain [23, 24, 25, 26, 27] or on a tree with fractal dimension less than two [28]. This algorithm builds from techniques [21] used to prove the area law in 1D ground states in these cases to produce an estimate for the ground state energy (and an MPS description of the ground state) to precision $\epsilon$ with probability at least $1 - \delta$ in

time $\text{poly}(L)\,\text{poly}(1/\epsilon)\,\text{poly}(\log(1/\delta))$; the algorithm has been called a "rigorous" version of DMRG (although it is quite different from DMRG in certain respects). (Unfortunately, optimizing the $\epsilon$-dependence of the algorithm has not been a focus of prior work).

**Heuristic quantum states for gapless systems**. As discussed above, the conjectured area law for critical quantum systems (e.g. ones with Fermi surfaces) contains logarithmic corrections. There are families of states that can be constructed to explicitly satisfy area laws with logarithmic corrections appropriate to critical states, e.g. the multiscale entanglement renormalization ansatz (MERA) in 1D [29], and the branching MERA in higher dimensions [30]. Because of the mildness of the logarithmic correction, it is also common to use explicit area law states, but with polynomially growing overhead, to model critical states (e.g. DMRG with $\text{poly}(L)$ bond dimension in 1D [31], or PEPS with $\text{poly}(L)$ bond dimension in 2D [32]).

## S1.5   Sufficient conditions for the validity of the adiabatic approximation

Let $s(t)$ be a "schedule" function such that $s(0) = 0$ and $s(T_{\text{ASP}}) = 1$, with $T_{\text{ASP}}$ the total time of the evolution. Let $\Upsilon(s(t))$ denote the state of the quantum computer at time $t$. The adiabatic-state-preparation (ASP) protocol calls for the initial state $\Upsilon(0)$ to be the ground state $\Upsilon_0$ of an initial Hamiltonian $H_{\text{init}}$, and for the state $\Upsilon(s)$ to subsequently evolve by the time-dependent Hamiltonian $H(s(t)) = (1 - s(t))H_{\text{init}} + s(t)H_{\text{final}}$ that interpolates from the initial Hamiltonian to the final Hamiltonian $H_{\text{final}}$. As $T_{\text{ASP}} \to \infty$, it is guaranteed that $|\langle \Phi | \Psi_0 \rangle| \to 1$, where $\Phi \equiv \Upsilon(1)$ is the end state of the ASP protocol and $\Psi_0 \equiv \Upsilon_0(1)$ is the ground state of $H_{\text{final}}$. For a desired success probability $p_0$ in QPE, one would like to choose the minimum value of $T_{\text{ASP}}$ such that $|\langle \Phi | \Psi_0 \rangle|^2 \geq p_0$; as discussed in SI section S2.5, the cost of QPE depends on the value of $p_0$ and taking $p_0 > 1/2$ is sufficient in our application.

Deducing rigorous analytical bounds on the total evolution time $T_{\text{ASP}}$ involves two main challenges. First, one must gain a sufficiently precise understanding of the eigenstructure of the

9

interpolating Hamiltonian $H(s)$; of particular importance is knowledge of a lower bound $\Delta$ on the ground-state gap that holds everywhere along the path. Finding such a bound is generally difficult, especially since situations where a numerical estimate of the ground state energy is most valuable are likely to be among the least amenable to analytical analysis. The numerics in the main text aim to compute such a bound, but this kind of numerical analysis will quickly become intractable as the system size increases. Second, even with complete knowledge of the spectral gap and other eigenstructure information, rigorous error bounds on $|\langle\Phi|\Psi_0\rangle|$ in the literature are quite complex. An often-quoted heuristic condition for ASP success is $T_{\text{ASP}} \gg \max_{s,i}|\langle\Upsilon_i(s)|\,dH/ds\,|\Upsilon_0(s)\rangle|/(E_i(s)-E_0(s))^2$, where $\Upsilon_i(s)$ and $E_i(s)$ are the instantaneous eigenstates and eigenvalues of $H(s)$ [33]. Usually, the maximum is achieved at $i=1$, in which case the above reads $T_{\text{ASP}} \gg \max_s \tau(s)$, motivating the "adiabatic time estimate" discussed in the main text. A weaker condition that can be rigorously proven for a large class of Hamiltonians is $T_{\text{ASP}} > K(\max_s\|dH/ds\|)\Delta^{-2}\log(\Delta^{-1})^{12}$ for some constant $K$, where $\|\cdot\|$ denotes the operator norm [34]; however, this bound is likely far from tight [35, 36].

These issues emphasize how, in most circumstances, it is expected that ASP will be applied heuristically, perhaps by trying increasingly larger values of $T_{\text{ASP}}$, as well as different choices of $H_{\text{init}}$ and schedule function $s(t)$, until success is observed. For a particular choice of $H_{\text{init}}$ and schedule, if the minimum eigenvalue gap $\Delta$ is exponentially small in $L$, ASP will fail for any choice of $T_{\text{ASP}} = \text{poly}(L)$.

On the other hand, the theoretical guarantees are strong enough to imply that $\Delta \geq 1/\text{poly}(L)$ is sufficient to guarantee that ASP succeeds in time $\text{poly}(L)$. Thus, as EQA is an asymptotic proposition, whether EQA is typically achievable via ASP + QPE boils down simply to whether $\Delta \geq 1/\text{poly}(L)$ and no efficient classical algorithm exists.

# S2  Supporting discussion and data for numerical experiments

## S2.1  Additional ansatz state preparation data

As a rough model for what happens when we try to improve the local fidelity for a large cluster such as FeMo-co, we performed DMRG calculations for the [2Fe-2S] cluster with two different oxidation states. We gradually increased the bond dimensions $D$ from 1 to 2000 and we computed the infidelity and the energy error with respect to an almost exact DMRG state with a large bond dimension of 8000 as $I = 1 - |\langle \Psi_0(D)|\Psi_0(8000)\rangle|^2$ and $\epsilon = E(D) - E(8000)$.

Fig. S1 shows the computational effort ($T$ defined in SI section S3.5) versus the inverse infidelity ($1/I$). The sublinear growth in the log-log scale means that the cost is poly($1/I$).

Similarly, Fig. S2 shows that the energy error is poly($I$). Thus, improving the local fidelity efficiently improves the local energy. Assuming some product like structure in the global state, this form of the local improvement leads to a global overlap consistent with a good energy.

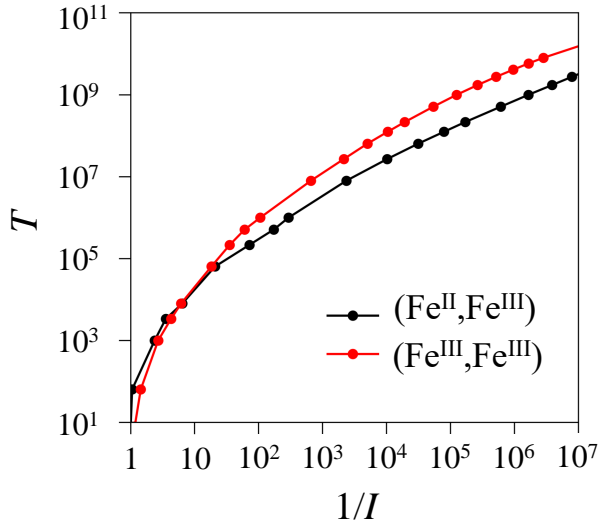

Figure S1: Computational effort ($T$ defined in SI section S3.5) with respect to the inverse infidelity ($1/I$).

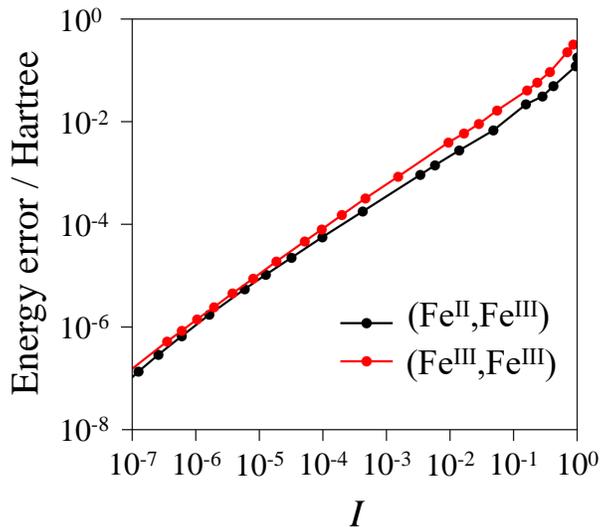

Figure S2: Energy error versus the infidelity ($I$).

## S2.2 Scaling of DMRG bond dimension in the Fe-S clusters

In the numerical study of FeS clusters, we performed DMRG calculations across a range of clusters (from [2Fe-2S] to the FeMo-co and P-clusters). Here we report on the DMRG bond dimension (in some cases, an estimate based on extrapolation) to reach constant $\bar{\epsilon}$ (here chosen as $\bar{\epsilon} = 10^{-3}$ Hartree per metal center). Fig. S3 shows the estimated DMRG bond dimension needed to reach this constant $\bar{\epsilon}$ as a function of the number of metal centers (2-8); the number of metals may be taken to be roughly proportional to $L$. Although there is some variation in the required bond dimension across clusters of the same nominal size (due to substantive differences in chemical composition, orbitals, and chemical structure amongst clusters with the same number of metals) we nonetheless can see that the bond dimension required for the 8Fe $P^N$ and $P^{ox}$ clusters is similar to that required for the smaller 4Fe clusters, and overall, there is a drop in slope between 2-4 and 4-8 metal ions. This reflects the relative similarity in cross-section when moving from 4 to 8 metal ions (and dissimilarity in cross-section when moving from 2 to 4 metal centers) and hints at a corresponding sub-volume (or area-law) behaviour of

12

the entanglement, as might be expected in a chemical ground-state problem.

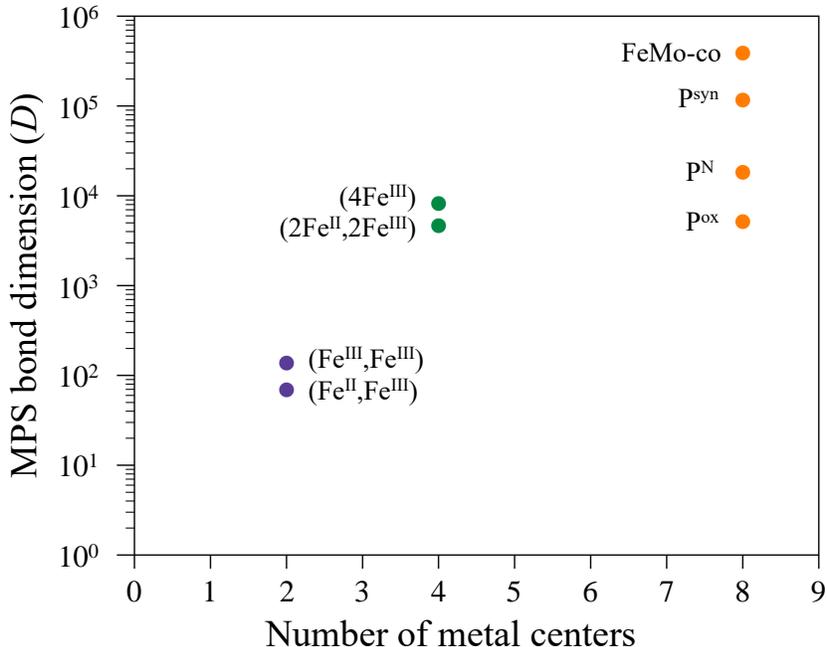

Figure S3: Required bond dimension ($D$) of the spin-adapted MPS for an estimated energy error of $10^{-3}$ Hartree per metal center, with respect to the number of metal centers.

## S2.3  Additional discussion of adiabatic state preparation data

In the main text we showed estimated adiabatic state preparation times for a variety of different mean-field Hamiltonians and interacting Hamiltonians. Here we provide some additional interpretation of the associated figures. The technical definition of the mean-field Hamiltonians and interacting Hamiltonians is provided in SI section S4.

As discussed in the main text, the lowest energy mean-field (Slater determinant) state - the lowest eigenstate of a Kohn-Sham mean-field Hamiltonian - does not have good overlap with the desired final state, and this is correlated with a large $T_{\text{ASP}}^{\text{est}}$. The set of mean-field data corresponds to considering initial $H(0)$ whose ground-state is an excited Slater determinant (relative to the lowest energy Slater determinant). In practice, this can be done by adding a

13

shift to the orbital energies to turn some excited-state orbitals into ground-state orbitals (see SI section S4). We consider over 100 such excited Slater determinants as initial states for the ASP procedure. Across these initial states, $T_{\text{ASP}}^{\text{est}}$ has a good correlation with the inverse weight $|\langle \Upsilon_0(0)|\Psi_0\rangle|^2$.

In addition, an entirely different set of initial states were obtained by choosing initial interacting Hamiltonians defined in active spaces $n_{\text{act}} < 12$ (definition in SI section S4). (For $n_{\text{act}} = 12$, determining the initial state would correspond to finding the ground-state of the full problem). However, the associated $T_{\text{ASP}}^{\text{est}}$ show the same inverse weight behaviour as those from the mean-field initial Hamiltonians, and the data from the two show a good correlation on the same plot. This suggests that in this problem the adiabatic state preparation time correlates well with the inverse initial weight, at least for this choice of adiabatic schedule.

We also see a linear trend when plotting $1/(\min_s \Delta(s))$ against $|\langle \Upsilon_0|\Psi_0\rangle|^2$ in Fig. S4. This shows that the minimum ground-state gap $(\min_s \Delta(s))$ is the main factor governing the behaviour of the adiabatic estimates.

## S2.4 Time-dependent simulations of adiabatic state preparation

When discussing the relationship between $T_{\text{ASP}}$ (for 75% final weight) and its adiabatic estimate $T_{\text{ASP}}^{\text{est}}$ in the main text, we concluded that ratio was $O(1)$, based on 17 time-dependent simulations of ASP shown in Fig. 2**B** of the main text. For these ASP simulations, we used 17 mean-field initial Hamiltonians and the linear-interpolated adiabatic path (definition in SI section S4). The weights of the 17 initial states ($\Upsilon_0(0)$) in the desired final ground state ($\Psi_0$) ranged from $4 \times 10^{-9}$ to $3 \times 10^{-2}$.

Here we show one example of these time-dependent ASP simulations, corresponding to a mean-field initial ground state ($\Upsilon_0$) with weight $|\langle \Upsilon_0(0)|\Psi_0\rangle|^2 = 2 \times 10^{-4}$. The adiabatic time estimate ($T_{\text{ASP}}^{\text{est}}$) for this initial state is $660\ E_h^{-1}$. We numerically propagated the wavefunction

14

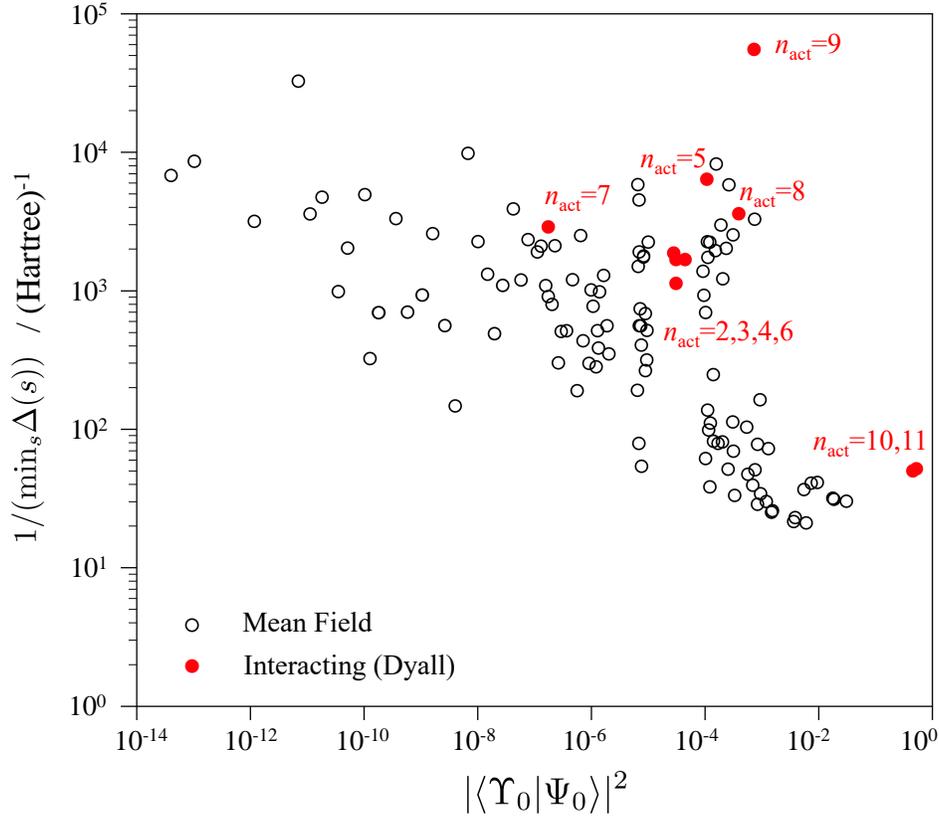

Figure S4: Inverse minimum ground-state gap ($1/(\min_s \Delta(s))$) along the adiabatic preparation path versus the weight of the initial ground state ($\Upsilon_0$) in the final ground state. ($\Psi_0$)

with a time-step $\Delta t = 0.01 E_h^{-1}$, via

$$|\Upsilon(s(t+\Delta t))\rangle = e^{-iH(s(t+\Delta t/2))\Delta t}|\Upsilon(s(t))\rangle. \tag{S1}$$

The action of the exponential was approximated by the Runge-Kutta 4th order method. Fig. S5 shows the weight of the adiabatically prepared state ($\Upsilon(s)$) as a function of the "schedule" function $s(t)$ (where $s(0) = 0$ and $s(T_{\text{ASP}}) = 1$). In this example, 75% final weight is achieved with $T_{\text{ASP}}/T_{\text{ASP}}^{\text{est}} \approx 2$.

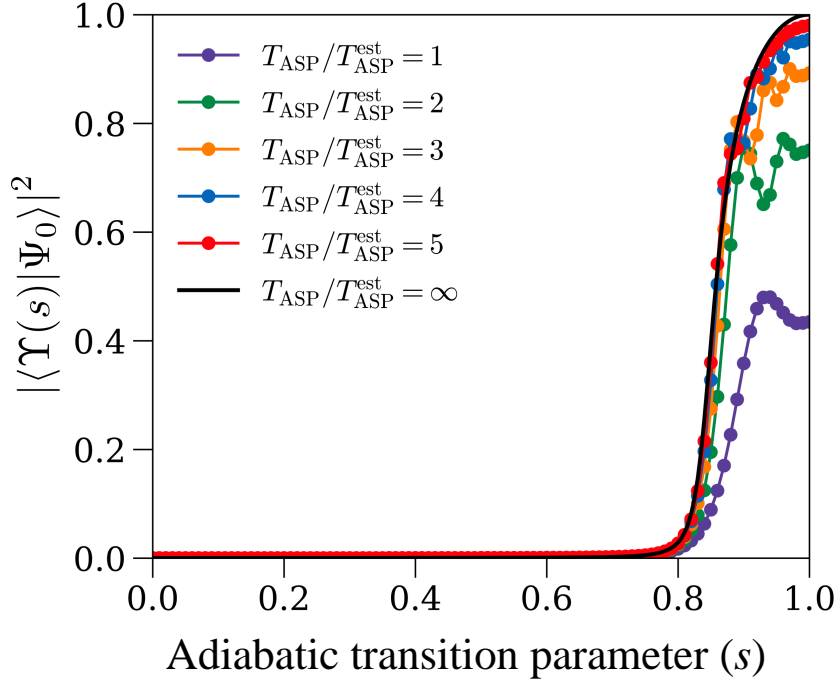

Figure S5: Weight $|\langle \Upsilon(s)|\Psi_0\rangle|^2$ of the adiabatically prepared state as a function of $s$ for different total ASP time $T_{\rm ASP}$ in time-dependent ASP simulations of a [2Fe-2S] model.

## S2.5 Evolution time for quantum phase estimation

We now describe the method we use for quantifying the evolution time required in performing QPE. By evolution time we mean the sum of all time $t$'s in the unitary operators of the form $e^{-iHt}$ that we need to implement. We assume we have an initial state $|\Phi\rangle$ satisfying $|\langle\Phi|\Psi_0\rangle|^2 = p_0 > 1/2$. We want to estimate the ground state energy to within error $\epsilon$ with probability at least 0.90. In a single run of QPE, if we want the success probability to be at least $p_s$ ($p_s < p_0$ because we are only considering the success probability of a single run), then according to the analysis in Section 5.2.1 in [37], the required evolution time is

$$T = \frac{\pi}{\epsilon(1 - p_s/p_0)}. \tag{S2}$$

16

Here by success we mean that the estimate is $\epsilon$-close to the ground state energy. Now we run QPE multiple times and take the median to boost the success probability. If more than half of the trials are successful, then the median will be $\epsilon$-close to the ground state energy. Therefore the boosted success probability is at least

$$p_{\text{boosted}} = \sum_{m=(n_{\text{trial}}+1)/2}^{n_{\text{trial}}} \binom{n_{\text{trial}}}{m} p_s^m (1-p_s)^{n_{\text{trial}}-m}. \tag{S3}$$

We then choose the minimal $n_{\text{trial}}$ that ensures $p_{\text{boosted}} \geq 0.90$. The evolution time required for QPE is then

$$n_{\text{trial}} T = \frac{\pi n_{\text{trial}}}{\epsilon(1 - p_s/p_0)}. \tag{S4}$$

In the above method, $p_s$ can be optimized to minimize the required evolution time. Also here we have only considered the simplest version of QPE, where the energy register is initialized to be the equal superposition of all bit strings. A better initialization using a resource state that minimizes the variance [38, 39], or the Kaiser window [40, 41], should help reduce the required evolution time. In the main text, a contrast is drawn between the QPE evolution time and the time for ASP; thus any further improvement in the QPE evolution time only serves to more strongly highlight the time cost of ASP.

## S2.6 Overview of systems chosen for scaling of classical heuristics

In the discussion of scaling of classical heuristics in the main text, we carried out illustrative calculations on large molecular problems (alkane chains and a protein fragment) as well as lattice problems relevant to correlated materials. Here we supply further background discussion regarding the relevance of these problems.

The molecular systems are representative of molecules encountered in organic and biological chemistry. The defining characteristic is the "single-reference" nature of the ground-state, where there is assumed to be good local overlap with a mean-field Slater determinant (although

17

the global overlap still decays like $e^{-\alpha L}$). In this setting, classical heuristics such as the coupled cluster method (see SI section S5) are often successful. The single-reference character is usually lost at points in a reaction where bonds are being broken or being formed. However, the non-trivial behaviour in the ground-state is then expected to be confined to a small number of atoms in an "active" region where the reaction is occurring. This is the basis of active space models, as for example, used to study the Fe-S metalloclusters embedded in the larger nitrogenase enzyme.

The second set of model problems is representative of models of correlated materials. The Heisenberg model is a model of a quantum magnet. On unfrustrated lattices, such as the cubic lattice, one can obtain exact results (up to sampling error) using quantum Monte Carlo due to the special sign-free structure of the Hamiltonian. This was used to generate exact data for this problem (see SI section S6.3). However, the tensor network calculations we presented on the same model do not take advantage of the sign-free structure; and further, the Heisenberg model is gapless, while the particular PEPS tensor network is designed to capture area law states. From this perspective, the 3D Heisenberg model might be expected to be hard for tensor network heuristics. However, as seen from the data, calculations on up to 1000 sites can be done with close to $O(L)$ cost for $\bar{\epsilon} \sim O(10^{-3})$. More interesting physics can be seen in quantum magnets on frustrated lattices or with frustrating interactions. Although larger bond dimensions are typically required to reach the same accuracy as in unfrustrated models, tensor network calculations on frustrated models can also reach sufficient accuracy to resolve the interesting physics, with poly($L$) cost, see e.g. [42, 43].

The 2D Hubbard model is a paradigmatic model of correlated materials. Similarly to as in the 2D Heisenberg model, the half-filled point in the phase diagram can be simulated by sign-free quantum Monte Carlo methods. However, away from half-filling the phase diagram is challenging, with the 1/8 doping point considered one of the most challenging points. For

18

narrow lattices, numerically exact results can be generated by DMRG, and we used this to generate exact benchmark data for the $4 \times 8$ and $4 \times 16$ lattices. However, the PEPS calculations reported in the main text do not in practice encounter the same exponential scaling with width as DMRG and can thus be used on wider lattices. Note that the specific 2D Hubbard model calculations do not represent the largest or most accurate tensor network calculations performed in this model; instead they were performed in such a way that the energies could be directly compared to DMRG calculations so that the error scaling could be assessed. For a more detailed study of the 2D Hubbard model using tensor networks, including the physics of competing low energy states, see e.g. Ref. [44].

We also note that tensor networks are by no means the only technique that can successfully treat strongly correlated models. For some alternatives in frustrated spin models, see e.g. Ref. [45]; for some other techniques used in challenging parts of the 2D Hubbard model phase diagram, see e.g. Ref. [44, 46].

Some of the techniques used in strongly correlated lattice models have now been generalized to ab initio Hamiltonians. For some examples, see Ref. [13]. With respect to tensor networks as used in the main text, the density matrix renormalization group and tree tensor networks have a long history of use in ab initio quantum chemistry [6, 47, 48, 49, 50]. PEPS have not yet been used in ab initio quantum chemistry, but generalizations towards ab initio models (e.g. calculations for fermionic continuum problems [51] and using long-range interactions [52]) have started to appear. Note that the physical principle of locality exploited by tensor networks is central to other heuristics for strongly correlated problems, most notably quantum embedding methods, such as dynamical mean-field theory (DMFT) [53] and density matrix embedding theory (DMET) [54]. Rather than target an ansatz for the global quantum state, these methods seek to assemble the relevant physical observables (e.g. Green's functions and density matrices) from calculations on local subsystems. They can be routinely applied to ab initio quantum

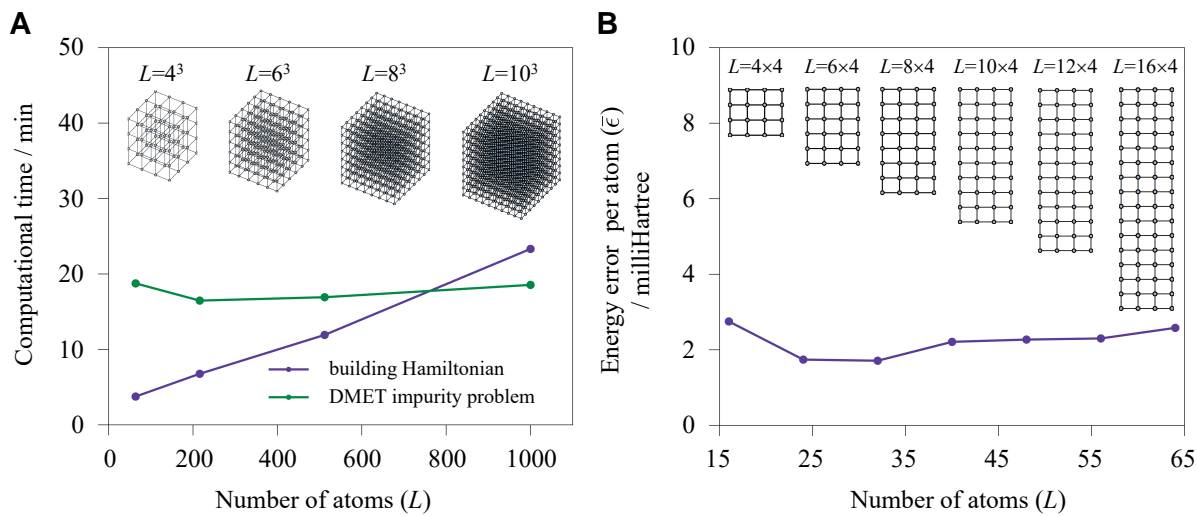

Figure S6: Ab initio calculations on hydrogen cubes and 2D lattices. The calculations are performed at a H-H separation of 1 Å. **A,** timing per atom for hydrogen cubes, **B,** error per atom for 2D lattices, compared against near-exact DMRG. **A,** using $2 \times 2 \times 2$ fragments, **B,** using $2 \times 2$ fragments.

chemical Hamiltonians [55, 56]. In Fig. S6**A** we show the timings for ab initio quantum chemistry DMET calculations on hydrogen cubes of size up to $10 \times 10 \times 10$ (minimal basis). This may be considered an ab initio analog of the Hubbard model. We see that the time per hydrogen atom increases only polynomially with system size. (Note that the time has two components: the time to solve the correlated DMET impurity problems, which is approximately constant per atom, and the time to set up the Hamiltonian for the DMET impurity problem for each atom, which grows polynomially with system size). In Fig. S6**B** we show DMET calculations with similar settings for a series of 2D hydrogen lattices of size $4 \times 4$ to $4 \times 16$, where we see that the absolute error achieved is close to 2 milliHartrees per atom. The precision scaling of embedding methods is currently not well characterized, but for some data and discussion of scaling of accuracy of local expectation values with embedding cluster size, see e.g. Refs. [57, 58]. These works provide numerical evidence of the convergence of local expectation values with cluster size (when appropriately evaluated) for a limited number of systems.

## S2.7 Error scaling and cost of local coupled cluster theory

In the main text, we showed that the computational effort to reach a given precision (error) is polynomial in the inverse precision. The computational time metric used was $L^{2m+2}$ (here $L$ is the number of orbitals), where $m$ is the excitation level ($= 2$ for CCSD, $3$ for CCSDT etc.). This is the cost scaling of the most expensive tensor contraction that dominates the computation in these methods.

In a local coupled cluster theory as shown in the main text, the time is reduced to $O(L)f(m)$, where $f(m)$ depends on the number of local orbitals $L_{\text{loc}}$ involved in the excitation, and is $\sim L_{\text{loc}}^{2m+2}$. Local versions of CC beyond CCSD(T) have not yet been implemented, however, assuming the same conjectured error scaling as without the local heuristic, the anticipated total cost (for problems where CC is a good heuristic) for local CC methods is $O(L)\text{poly}(1/\epsilon)$.

## S3 Technical details: electronic ground-state calculations for the Fe-S clusters

### S3.1 Active space models

In our Fe-S simulations, we used active space models which capture the important aspects of the electronic structure of these systems. In particular, we used active space models from previous studies for the [2Fe-2S] and [4Fe-4S] [59], P-cluster [60], and FeMo-cofactor [61] systems. Detailed information about these models can be found in Refs. [59, 60, 61]. In this subsection, we briefly summarize the approximations used to construct the electronic models.

Optimized BS-DFT geometries were used for the synthetic [2Fe-2S] and [4Fe-4S] clusters [62] while X-ray crystal structures were used for the P-clusters [63, 64] and the FeMo-cofactor [65]. All the active space models were constructed from split-localized [66, 67] natural orbitals from high-spin unrestricted DFT wavefunctions incorporating scalar relativistic effects. Cal-

culations on the [2Fe-2S] and [4Fe-4S] clusters used the BP86 functional [68, 69], TZP-DKH basis [70], and the spin-free exact two-component (sf-X2C) Hamiltonian [71, 72, 73]. Calculations on the P-clusters used the BP86 functional [68, 69], def2-SVP basis [70], and the sf-X2C Hamiltonian [71, 72, 73]. Calculations on the FeMo-co clusters used the B3LYP functional, [74, 75, 76] TZP-DKH basis [70] for Fe, S, and Mo, def2-SVP basis [70] for the other atoms (C, H, O, and N), and the sf-X2C Hamiltonian [71, 72, 73]. In addition, the calculations on the P-clusters and the FeMo-cofactor used the conductor like screening model (COSMO) [77] with a dielectric constant $\epsilon = 4.0$ to mimic the protein environment.

Finally, all the active space models were constructed from the full valence space of the Fe $3d$, S $3p$, and bonding ligand orbitals around each metal atom. The active space model for the FeMo-cofactor also contained the additional full valence space of the Mo $4d$ and central C $2s$ and $2p$. The resulting active space models consisted of CAS(30e,20o) for $(Fe^{III}, Fe^{III})$, CAS(31e,20o) for $(Fe^{II}, Fe^{III})$, CAS(54e,36o) for $(2 Fe^{II}, 2 Fe^{III})$, CAS(52e,36o) for $(4 Fe^{III})$, CAS(108e,71o) for $P^{syn}$, CAS(114e,73o) for $P^N$, CAS(120e,77o) for $P^{ox}$, and CAS(113e,76o) for the FeMo-cofactor.

## S3.2 Initial guesses for DMRG calculations

There are many local electronic minima in the energy landscape of ansatz approximations to the P-cluster and FeMo-cofactor ground-state wavefunctions. To reach the correct ground-state, it is important to start from a good initial guess. We used a special DMRG initialization procedure presented in earlier studies [59, 60]. In this subsection, we briefly summarize the procedure.

As an initial guess to the spin-adapted DMRG calculation, we first performed a spin-projected (SP)-MPS calculation [59]. This was initialized by spin-projecting a broken-symmetry determinant chosen to represent the spin-coupling pattern shown in Fig. S7 with the total spin, total charge, and the oxidation number states of the metal cluster as summarized in Table S1. The

22

resulting spin-projected MPS was further optimized up to a bond dimension of 50, and then fed to the spin adapted DMRG calculation.

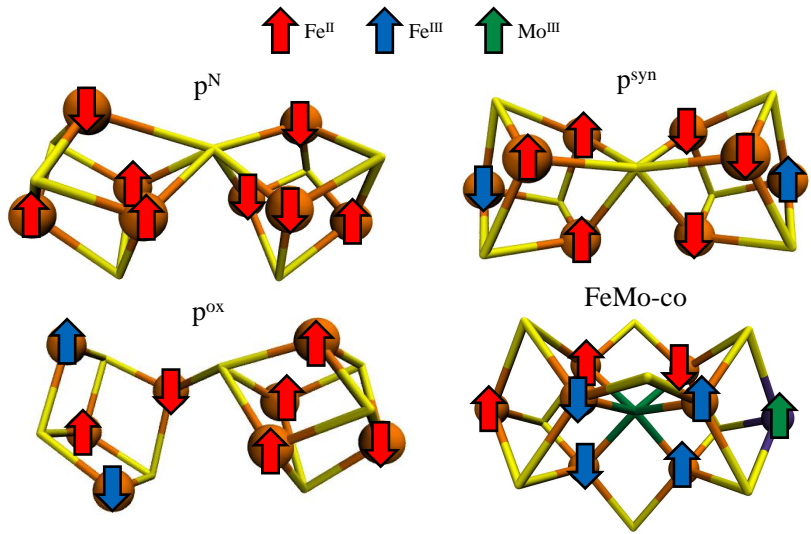

Figure S7: Spin-coupling pattern of the broken symmetry guess for the electronic ground states reported in previous theoretical studies [60, 78].

Table S1: Total spin and total charge of the metal clusters and the oxidation number of each metal atom for the broken symmetry guess of the electronic ground state.

|  | Total spin | Total charge | Oxidation number |
| --- | --- | --- | --- |
| $p^N$ | $S = 0$ | $-4$ | $4Fe^{II}\uparrow\ 4Fe^{II}\downarrow$ |
| $p^{syn}$ | $S = 0$ | $-2$ | $Fe^{III}\uparrow\ Fe^{III}\downarrow\ 3Fe^{II}\uparrow\ 3Fe^{II}\downarrow$ |
| $p^{ox}$ | $S = 4$ | $-4$ | $Fe^{III}\uparrow\ Fe^{III}\downarrow\ 4Fe^{II}\uparrow\ 2Fe^{II}\downarrow$ |
| FeMo-co | $S = 1.5$ | $-3$ | $2Fe^{III}\uparrow\ 2Fe^{III}\downarrow\ 2Fe^{II}\uparrow\ Fe^{II}\downarrow\ Mo^{III}\uparrow$ |

## S3.3 Extrapolations for the total DMRG energies

We extrapolated the ground-state DMRG energy as a function of discarded weight, see e.g. Ref. [79]. Except for FeMo-co, the MPS energies at smaller bond dimension were obtained from "reverse sweep schedule" DMRG calculations, starting from the converged MPS with the

23

largest bond dimension. We excluded the energy with the largest bond dimension, which was not fully converged, from the energy extrapolations.

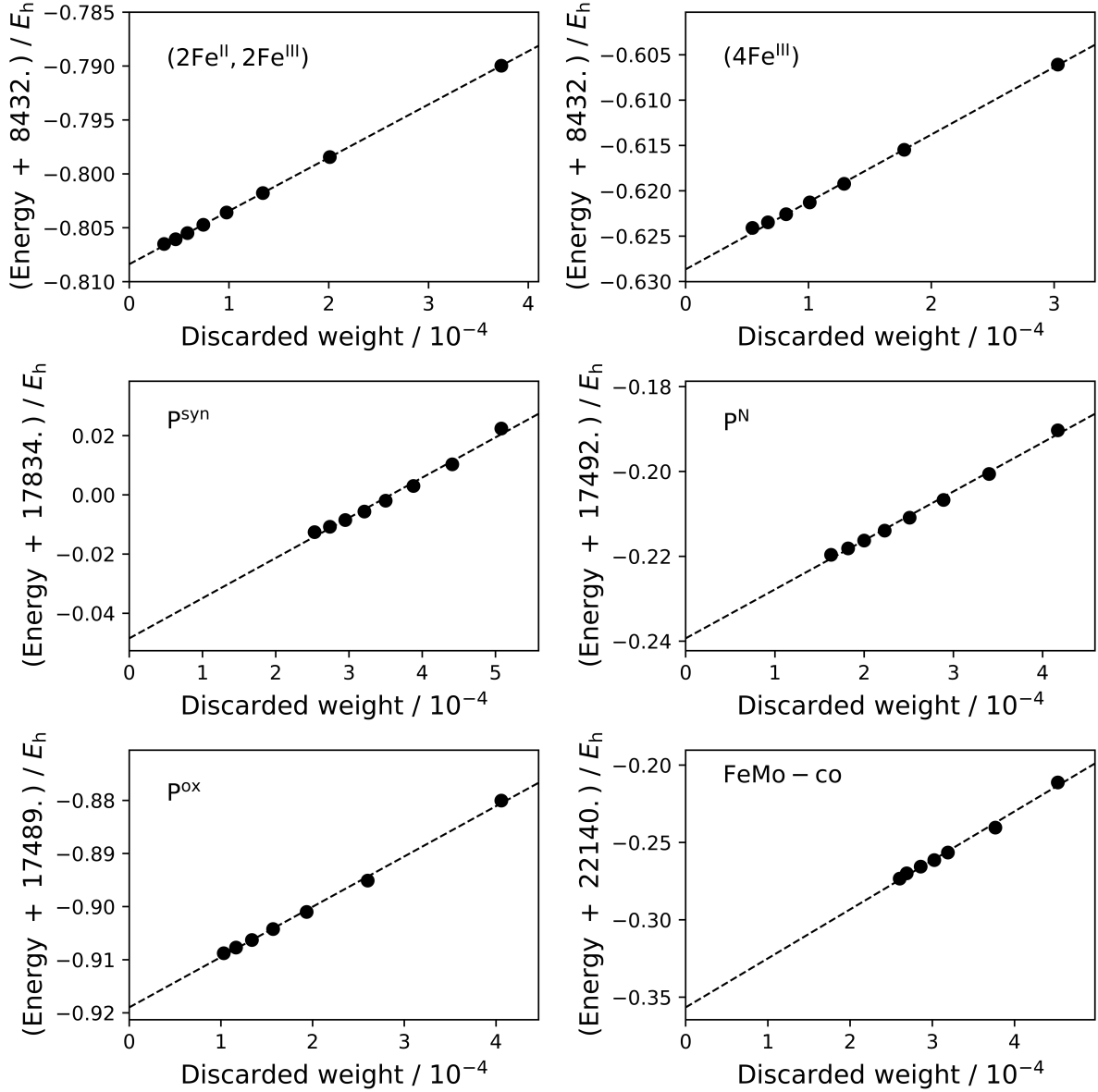

Figure S8: DMRG energy in $E_h$ of the Fe-S clusters versus the discarded weight. The dashed lines represent the best fit straight lines which are extrapolated to zero discarded weight to obtain an estimate of the exact energy.

The above extrapolations indicate that for the [4Fe-4S] clusters, the extrapolated exact en-

Table S2: Total DMRG energies in $E_h$ for the electronic ground states of the Fe-S clusters. $D$ denotes the bond dimension in the spin-adapted MPS ansatz.

| $D$ | Discarded weight | Energy / $E_h$ | $D$ | Discarded weight | Energy / $E_h$ |
|---|---|---|---|---|---|
| | 2Fe(II)-2Fe(III) | | | 4Fe(III) | |
| 1000 | $3.73 \times 10^{-4}$ | -8432.789966 | 1000 | $3.03 \times 10^{-4}$ | -8432.606085 |
| 2000 | $2.01 \times 10^{-4}$ | -8432.798444 | 2000 | $1.78 \times 10^{-4}$ | -8432.615484 |
| 3000 | $1.34 \times 10^{-4}$ | -8432.801780 | 3000 | $1.29 \times 10^{-4}$ | -8432.619229 |
| 4000 | $9.76 \times 10^{-5}$ | -8432.803590 | 4000 | $1.01 \times 10^{-4}$ | -8432.621279 |
| 5000 | $7.43 \times 10^{-5}$ | -8432.804722 | 5000 | $8.18 \times 10^{-5}$ | -8432.622578 |
| 6000 | $5.84 \times 10^{-5}$ | -8432.805500 | 6000 | $6.70 \times 10^{-5}$ | -8432.623476 |
| 7000 | $4.65 \times 10^{-5}$ | -8432.806069 | 7000 | $5.44 \times 10^{-5}$ | -8432.624094 |
| 8000 | $3.50 \times 10^{-5}$ | -8432.806502 | $\infty$ | | -8432.6287(9) |
| $\infty$ | | -8432.8084(4) | | | |
| | $P^{syn}$ | | | $P^{N}$ | |
| 1000 | $5.08 \times 10^{-4}$ | -17833.977588 | 1000 | $4.17 \times 10^{-4}$ | -17492.190311 |
| 1500 | $4.41 \times 10^{-4}$ | -17833.989684 | 1500 | $3.40 \times 10^{-4}$ | -17492.200594 |
| 2000 | $3.88 \times 10^{-4}$ | -17833.996979 | 2000 | $2.89 \times 10^{-4}$ | -17492.206725 |
| 2500 | $3.50 \times 10^{-4}$ | -17834.001975 | 2500 | $2.51 \times 10^{-4}$ | -17492.210894 |
| 3000 | $3.21 \times 10^{-4}$ | -17834.005658 | 3000 | $2.23 \times 10^{-4}$ | -17492.213953 |
| 3500 | $2.95 \times 10^{-4}$ | -17834.008503 | 3500 | $2.00 \times 10^{-4}$ | -17492.216294 |
| 4000 | $2.74 \times 10^{-4}$ | -17834.010764 | 4000 | $1.82 \times 10^{-4}$ | -17492.218146 |
| 4500 | $2.53 \times 10^{-4}$ | -17834.012610 | 4500 | $1.63 \times 10^{-4}$ | -17492.219644 |
| 5000 | $2.27 \times 10^{-4}$ | -17834.014051 | 5000 | $1.36 \times 10^{-4}$ | -17492.220847 |
| $\infty$ | | -17834.049(7) | $\infty$ | | -17492.239(4) |
| | $P^{ox}$ | | | FeMo-co | |
| 500 | $4.06 \times 10^{-4}$ | -17489.880043 | 2000 | $4.53 \times 10^{-4}$ | -22140.21120 |
| 1000 | $2.60 \times 10^{-4}$ | -17489.895112 | 3000 | $3.77 \times 10^{-4}$ | -22140.24039 |
| 1500 | $1.94 \times 10^{-4}$ | -17489.901003 | 4000 | $3.19 \times 10^{-4}$ | -22140.25652 |
| 2000 | $1.57 \times 10^{-4}$ | -17489.904229 | 4500 | $3.02 \times 10^{-4}$ | -22140.26141 |
| 2500 | $1.34 \times 10^{-4}$ | -17489.906285 | 5000 | $2.86 \times 10^{-4}$ | -22140.26562 |
| 3000 | $1.17 \times 10^{-4}$ | -17489.907715 | 5500 | $2.69 \times 10^{-4}$ | -22140.26995 |
| 3500 | $1.03 \times 10^{-4}$ | -17489.908781 | 6000 | $2.60 \times 10^{-4}$ | -22140.27335 |
| 4000 | $8.70 \times 10^{-5}$ | -17489.909590 | 6500 | $2.55 \times 10^{-4}$ | -22140.27633 |
| $\infty$ | | -17489.919(2) | 7000 | $2.47 \times 10^{-4}$ | -22140.27881 |
| | | | $\infty$ | | -22140.36(2) |

ergy is between 3-5 m$E_h$ (extrapolation distance) below the lowest DMRG variational energy, approximately 1 m$E_h$ per Fe. It is difficult to estimate the error of the extrapolated energy itself, but it can conservatively be taken to be some fraction of the extrapolation distance. (Note that the designation of chemical accuracy ($\sim$ 1kcal/mol) was originally intended to refer to thermochemical energy *differences*, e.g. heats of formation and ionization potentials, mainly in small molecules [80]). In the absence of a specific chemical transformation to discuss (which would involve changing the electronic structure of only a part of the system) the more relevant metric on the total energy is often $\bar{\epsilon}$ rather than $\epsilon$, because chemical reactions usually involve changes in a small part of the total system.

In the P- and FeMo-co clusters, the extrapolation distance is somewhat larger, ranging from about 10 m$E_h$ (P$^{ox}$) to more than 80 m$E_h$ in FeMo-co (approximately 1-10 m$E_h$ per metal atom), reflecting the greater complexity. Although the extrapolation distance in the FeMo-cofactor is larger than desired, the convergence of the energy remains superpolynomial in $1/\epsilon$ (see Fig. S9) as for the other clusters, and the extrapolation is sufficiently systematic to provide some confidence in the extrapolated energy.

## S3.4 Extracting the largest weight determinant and configuration state function

We extracted the (approximately) largest weight Slater determinant ($\Phi_D$) and configuration state function ($\Phi_{\text{CSF}}$) from the optimized MPS ($\Psi_0(D)$) with bond dimension $D$, by gradually compressing the MPS down to a bond dimension of one. This provides an approximate way to obtain a Slater determinant (from compressing a non-spin-adapted MPS) or configuration state function (CSF) (spin-adapted MPS) that has the largest weight in the MPS. (It should be noted that this procedure is not guaranteed to find the absolutely largest weight determinant or CSF, since the compression is done as a sweep rather in a globally optimal way.) The results are

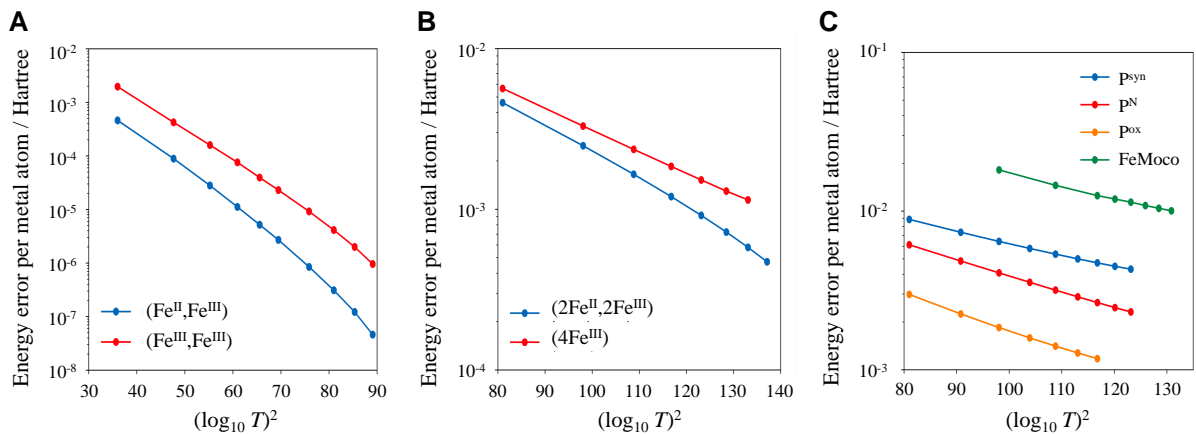

Figure S9: Convergence of energy error per metal atom of the Fe-S clusters with respect to computational effort ($T$ defined in SI section S3.5). The observed empirical $\log(1/\bar{\epsilon}) \sim (\log T)^2$ relation corresponds to superpolynomial $1/T$ error convergence. The errors are obtained with respect to the exact FCI energies for the [2Fe-2S] systems (**A**), and extrapolated DMRG energies in the other clusters. (**B** and **C**)

summarized in Table S3.

### S3.5 Computational time metric

The computational cost of the DMRG quantum chemistry algorithm is $O(D^3 L^3) + O(D^2 L^4)$. For the DMRG calculations of the Fe-S clusters, the bond dimension (c.a. $> 1000$) is much larger than the number of active space orbitals (c.a. $< 100$), thus the cubic term dominates. The metric of DMRG computational time is $T = D^3$ in Fig. 2 of the main text; this factorizes out the system size dependence of the Hamiltonian, allowing for a simpler comparison between the different systems for quantities controlled by $D$ (such as the error). Fig. S10 shows the strong linear relation between this time metric and the actual CPU computational time in seconds.

### S3.6 Additional convergence data for the [2Fe-2S] clusters

As the exact FCI data is available for the [2Fe-2S] clusters, in Fig. S11 we show some additional plots of DMRG convergence with bond dimension, for various quantities.

27

Table S3: Overlap of the largest weight Slater determinant ($|\langle\Phi_D|\Psi_0\rangle|$) and overlap of the largest weight configuration state function ($|\langle\Phi_{\text{CSF}}|\Psi_0\rangle|$) in the MPS wavefunction with bond dimension $D$; the weights plotted in the main text correspond to the squared overlap. The $D = \infty$ overlap uses an empirical extrapolation form $S(D) = c/D + S(\infty)$.

| $D$ | Overlap ($S$) CSF | Overlap ($S$) DET | $D$ | Overlap ($S$) CSF | Overlap ($S$) DET |
|---|---|---|---|---|---|
| | 2Fe(II)-2Fe(III) | | | 4Fe(III) | |
| 2000 | $1.66 \times 10^{-2}$ | $3.54 \times 10^{-3}$ | 2000 | $7.12 \times 10^{-2}$ | $2.15 \times 10^{-2}$ |
| 3000 | $1.61 \times 10^{-2}$ | $3.44 \times 10^{-3}$ | 3000 | $6.89 \times 10^{-2}$ | $2.08 \times 10^{-2}$ |
| 4000 | $1.58 \times 10^{-2}$ | $4.99 \times 10^{-3}$ | 4000 | $6.75 \times 10^{-2}$ | $2.03 \times 10^{-2}$ |
| 5000 | $1.56 \times 10^{-2}$ | $4.93 \times 10^{-3}$ | 5000 | $6.65 \times 10^{-2}$ | $2.01 \times 10^{-2}$ |
| 6000 | $1.54 \times 10^{-2}$ | $4.88 \times 10^{-3}$ | 6000 | $6.59 \times 10^{-2}$ | $1.99 \times 10^{-2}$ |
| 7000 | $1.53 \times 10^{-2}$ | $4.84 \times 10^{-3}$ | 7000 | $6.54 \times 10^{-2}$ | $1.97 \times 10^{-2}$ |
| $\infty$ | $1.47 \times 10^{-2}$ | $4.65 \times 10^{-3}$ | $\infty$ | $6.34 \times 10^{-2}$ | $1.90 \times 10^{-2}$ |
| | P$^{\text{syn}}$ | | | P$^{\text{N}}$ | |
| 1000 | $4.51 \times 10^{-3}$ | $1.97 \times 10^{-3}$ | 1000 | $2.44 \times 10^{-2}$ | $1.10 \times 10^{-2}$ |
| 1500 | $4.32 \times 10^{-3}$ | $2.05 \times 10^{-3}$ | 1500 | $2.05 \times 10^{-2}$ | $1.03 \times 10^{-2}$ |
| 2000 | $4.09 \times 10^{-3}$ | $2.04 \times 10^{-3}$ | 2000 | $1.80 \times 10^{-2}$ | $9.73 \times 10^{-3}$ |
| 2500 | $3.93 \times 10^{-3}$ | $2.02 \times 10^{-3}$ | 2500 | $1.62 \times 10^{-2}$ | $9.22 \times 10^{-3}$ |
| 3000 | $3.80 \times 10^{-3}$ | $2.00 \times 10^{-3}$ | 3000 | $1.49 \times 10^{-2}$ | $8.81 \times 10^{-3}$ |
| 3500 | $3.69 \times 10^{-3}$ | $1.99 \times 10^{-3}$ | 3500 | $1.39 \times 10^{-2}$ | $8.48 \times 10^{-3}$ |
| 4000 | $3.59 \times 10^{-3}$ | $1.98 \times 10^{-3}$ | 4000 | $1.32 \times 10^{-2}$ | $8.24 \times 10^{-3}$ |
| 4500 | $3.51 \times 10^{-3}$ | $1.97 \times 10^{-3}$ | 4500 | $1.27 \times 10^{-2}$ | $8.07 \times 10^{-3}$ |
| 5000 | $3.45 \times 10^{-3}$ | $1.96 \times 10^{-3}$ | 5000 | $1.23 \times 10^{-2}$ | $7.95 \times 10^{-3}$ |
| $\infty$ | $3.15 \times 10^{-3}$ | $1.92 \times 10^{-3}$ | $\infty$ | $8.63 \times 10^{-3}$ | $6.83 \times 10^{-3}$ |
| | P$^{\text{ox}}$ | | | FeMo-co | |
| 500 | $5.66 \times 10^{-3}$ | $1.94 \times 10^{-3}$ | 6000 | $3.68 \times 10^{-4}$ | |
| 1000 | $5.33 \times 10^{-3}$ | $1.78 \times 10^{-3}$ | 7000 | | $9.88 \times 10^{-5}$ |
| 1500 | $5.33 \times 10^{-3}$ | $1.70 \times 10^{-3}$ | | | |
| 2000 | $5.08 \times 10^{-3}$ | $1.65 \times 10^{-3}$ | | | |
| 2500 | $5.00 \times 10^{-3}$ | $1.61 \times 10^{-3}$ | | | |
| 3000 | $4.94 \times 10^{-3}$ | $1.57 \times 10^{-3}$ | | | |
| 3500 | $4.90 \times 10^{-3}$ | $1.55 \times 10^{-3}$ | | | |
| 4000 | $4.86 \times 10^{-3}$ | $1.53 \times 10^{-3}$ | | | |
| $\infty$ | $4.72 \times 10^{-3}$ | $1.44 \times 10^{-3}$ | | | |

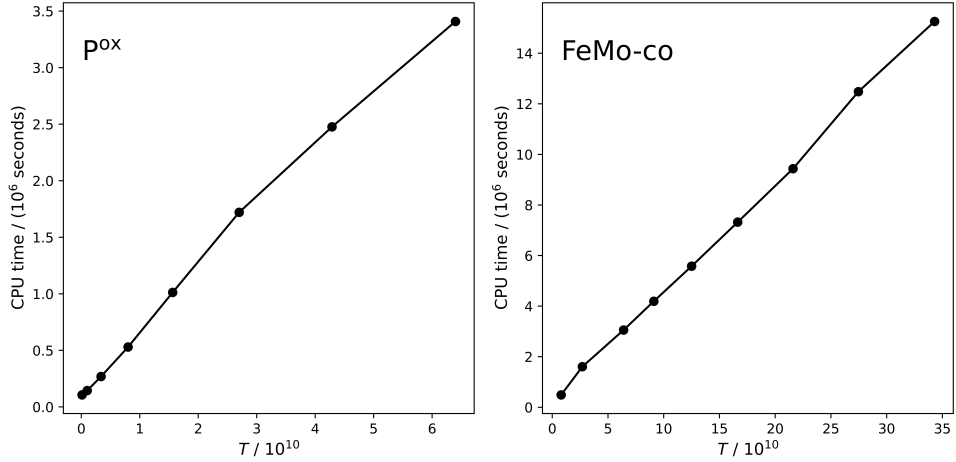

Figure S10: Linear relation between the DMRG time metric used in the main text ($T = D^3$) and the actual computational time (seconds) for a DMRG sweep with the bond dimension $D$.

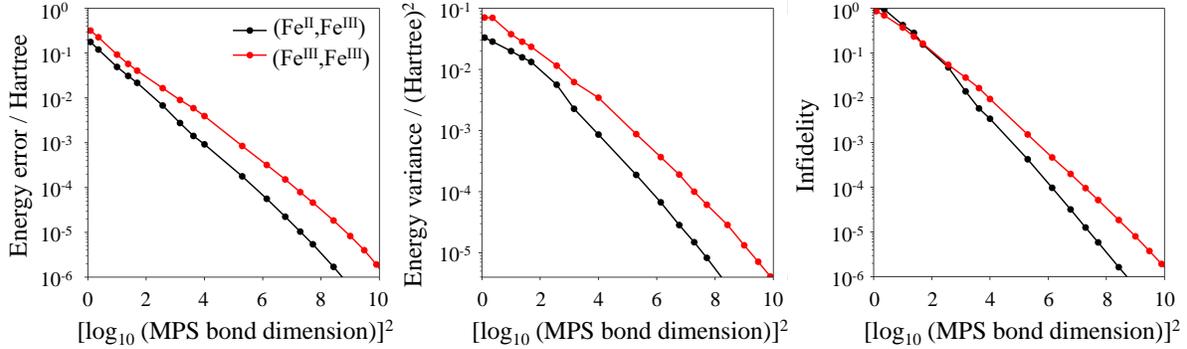

Figure S11: Convergence of the energy error ($E_{\text{DMRG}} - E_{\text{FCI}}$), the energy variance ($\langle \Phi_{\text{DMRG}} | (H - E_{\text{DMRG}})^2 | \Phi_{\text{DMRG}} \rangle$), and the infidelity ($1 - |\langle \Phi_{\text{DMRG}} | \Psi_{\text{FCI}} \rangle|^2$) with the MPS bond dimension ($D$) for the [2Fe-2S] models.

## S4 Technical details: Adiabatic state preparation

### S4.1 CAS model for adiabatic state preparation

We examined ASP for the singlet ground state of the [2Fe-2S] cluster, ($Fe^{III}$, $Fe^{III}$). Because the corresponding classical simulations are very expensive (due to the need to prepare states very slowly) we constructed smaller active space models. We used a Kohn-Sham orbital active space derived from DFT calculations using the BP86 functional and the scalar relativistic X2C

Hamiltonian using the TZP-DKH basis. From these calculations, we obtained a CAS(14e,12o) model and verified that it produced similar chemistry to that of the CAS(30e,20o) model used in the FCI calculations. The 12 orbitals consist of the ten $3d$ orbitals of the irons and two out-of-plane $3p$ orbitals of the bridging sulfurs; "similar chemistry" here means that the Hamiltonian yields the same order of low-lying spin states, similar shapes of the natural orbitals and similar natural orbital occupations.

The Hamiltonian of this CAS(14e,12o) model can be written as

$$H = \sum_{pq} h_{pq}^{\mathrm{C}} \hat{E}_{pq} + \frac{1}{2} \sum_{pqrs} (pq|rs)(\hat{E}_{pq}\hat{E}_{rs} - \delta_{qr}\hat{E}_{ps}) + K_{\mathrm{C}}, \tag{S5}$$

with

$$\hat{E}_{pq} = \hat{a}_{p\alpha}^{\dagger}\hat{a}_{q\alpha} + \hat{a}_{p\beta}^{\dagger}\hat{a}_{q\beta}, \tag{S6}$$

$$h_{pq}^{\mathrm{C}} = h_{pq} + \sum_{I}(2(pq|II) - (pI|Iq)), \tag{S7}$$

$$K_{\mathrm{C}} = 2\sum_{I} h_{II} + \sum_{IJ}(2(II|JJ) - (IJ|JI)), \tag{S8}$$

where the orbital indices $\{p, q, r, s\}$ are used for the 12 orbitals in the active space, $\{I, J\}$ are used for orbitals in the core (doubly occupied) space, and $\alpha, \beta$ are the spin indices. $h_{pq}$ and $(pq|rs)$ are the one-electron and two-electron integrals, respectively; $h_{pq}^{\mathrm{C}}$ is the effective one-electron integral (that includes the interactions with the core orbitals) and $K_{\mathrm{C}}$ is the core energy.

## S4.2 Initial Hamiltonian for the adiabatic state preparation

The ASP protocol prepares the state $\Phi_0$ by starting in the ground state of an easy-to-prepare Hamiltonian. We tested two types of initial Hamiltonians.

**Mean-field Hamiltonian:** We prepared a mean-field initial Hamiltonian defined as the sum of the Kohn-Sham Fock operators. In addition, we added an energy shift ($\delta$) to target a specific mean-field determinant as the ground state and to introduce a non-trivial energy gap between

30

the ground and first excited state. Using the same notation for the orbital indices introduced in Eq. (S5), the mean-field Hamiltonian is

$$H_{\text{Fock}} = \sum_p (\epsilon_p + \delta(n_p))\hat{E}_{pp} + K_{\text{C,Fock}}, \tag{S9}$$

where $\epsilon_p$ is the Kohn-Sham orbital energy of the $p$th orbital and $K_{\text{C,Fock}} = 2\sum_I \epsilon_I$ is the energy from the core orbitals. The shift $\delta$ is a function of the occupation number ($n_p$) of the $p$th orbital ($n_p$ corresponds to the occupations in the targeted mean-field determinant). We used $\delta(n_p = 2) = 0$ and $\delta(n_p = 0) = 0.5$ (Hartree). We find that the energy shift ensures that the targeted mean-field state is the ground state and the initial energy gap is larger than $0.35$ Hartree.

**Interacting Hamiltonian:** We chose as our interacting Hamiltonian the Dyall Hamiltonian, [81], the zeroth-order Hamiltonian of the $n$-electron valence state perturbation theory (NEVPT) [82]. To generate different interacting Hamiltonians with active space sizes $n_{\text{act}} \leq 12$, we further divided the CAS(14e, 12o) model into inactive, "inner active", and virtual spaces, with indices $\{i, j, k, l\}$, $\{u, v, w, x\}$, and $\{a, b, c, d\}$, respectively. We always included the two S $3p$ orbitals in the inner active space, and added the Fe $3d$ orbitals around the highest occupied molecular orbital to generate larger inner active spaces.

The Dyall Hamiltonian uses effective Fock operators in the inactive and virtual spaces, and the full two-body Hamiltonian in the (inner) active space.

$$\begin{aligned} H_{\text{Dyall}} = &\sum_{ij} f_{ij}\hat{E}_{ij} + \sum_{ab} f_{ab}\hat{E}_{ab} \\ &+ \sum_{uv} h^{\text{eff}}_{uv}\hat{E}_{uv} + \frac{1}{2}\sum_{uvwx}(uv|wx)(\hat{E}_{uv}\hat{E}_{wx} - \delta_{vw}\hat{E}_{ux}) + K_{\text{C,Dyall}}, \end{aligned} \tag{S10}$$

with

$$h_{pq}^{\text{eff}} = h_{pq}^{\text{C}} + \sum_{i}(2(pq|ii) - (pi|iq)), \tag{S11}$$

$$f_{pq} = h_{pq}^{\text{eff}} + \sum_{uv} D_{uv}((pq|uv) - 0.5(pu|vq)), \tag{S12}$$

$$K_{\text{C,Dyall}} = K_{\text{C}} + 2\sum_{i} h_{ii}^{\text{C}} + \sum_{ij}(2(ii|jj) - (ij|ji)) - 2\sum_{i} f_{ii}. \tag{S13}$$

Here, $h_{pq}^{\text{C}}$ is the effective one-electron integral including the interactions with the core orbitals defined in Eq. (S7) and $D_{uv}$ is the spin-traced one-particle density matrix in the (inner) active space. We obtained this density matrix by diagonalizing the (inner) active space problem first.

## S5 Local coupled cluster calculations

The coupled-cluster ansatz obtains the exact state $|\Psi_0\rangle$ from a trial state $|\Phi_0\rangle$ by the action of a wave operator encoded as an exponential of a Fock-space *cluster* operator $\hat{T}$:

$$|\Psi_0\rangle = \exp(\hat{T})|\Phi_0\rangle; \tag{S14}$$

in practice $|\Phi_0\rangle$ is almost always a single (Hartree-Fock) determinant. The cluster operator is usually written as a sum of (particle-hole) excitations $\hat{T} = \hat{T}_1 + \hat{T}_2 + \hat{T}_3 \ldots$, where $\hat{T}_1$ creates single excitations, $\hat{T}_2$ creates double excitations, etc. Note that for finite truncations of this sum, the exponential still ensures that excitations of large numbers of particles and holes are created, and one can have $\Psi_0$ such that as $L \to \infty$, $\langle \Phi_0|\Psi_0\rangle = 0$. $\Psi_0$ which can be accurately represented by (low-order) truncations of $\hat{T}$ are typically called single-reference problems.

For quantitative accuracy in such single-reference chemicaal systems, empirical studies show that $\hat{T}$ must include at least 1- and 2-body excitation operators ("coupled-cluster singles and doubles", CCSD) augmented by an approximate treatment of 3-body excitations, such as in the CCSD(T) method[83, 84]. Higher order excitations, including quadruples, quintuples, etc., can also be included for small molecules, as shown by the data in the main text [85].

The local CCSD(T) method (DLPNO-CCSD(T)) illustrated in the main text [86, 87] defines the cluster operator excitations from localized orbitals occupied in $|\Phi_0\rangle$ to orbital- and orbital-pair-specific unoccupied orbitals; this allows to encode the cluster operator with $O(L^2)$ terms. Further pruning of the $\hat{T}$ operator to short-range 2- and 3-body clusters results in the linear size complexity of the cluster operator and linear complexity of its solver. To reduce the discretization error, the cluster operator is augmented by explicitly correlated[1] 2-body excitations[87].

## S5.1 The coupled-cluster-based protocol for the enthalpy of formation of $n$-alkanes

The explicitly correlated local CCSD(T) (DLPNO-CCSD(T)-F12[87]) method was used to compute the standard enthalpies of formation of $n$-alkanes at 298.15 K largely following the procedure benchmarked in Ref. 88. The optimal geometries for $n$-alkanes were determined at the Kohn-Sham DFT (B3LYP/pc-2) level of theory. The electronic atomization energy was computed as

$$\Delta E_{\text{CC}}(\text{C}_m\text{H}_{2m+2}) = mE_{\text{CC}}(\text{C}) + (2m+2)E_{\text{HF}}(\text{H}) - E_{\text{CC}}(\text{C}_m\text{H}_{m+2}), \qquad \text{(S15)}$$

where $E_{\text{HF}}$ and $E_{\text{CC}}$ denote the Hartree-Fock/cc-pVTZ-F12 and the coupled-cluster electronic energies, respectively. The latter was evaluated as

$$E_{\text{CC}} = E_{\text{DLPNO-CCSD(T)-F12(fc)/cc-pVTZ-F12}} + \left(E_{\text{DLPNO-CCSD(T)(ae)/cc-pCVTZ}} - E_{\text{DLPNO-CCSD(T)(fc)/cc-pCVTZ}}\right) \qquad \text{(S16)}$$

where "fc"/"ae" denote the computations with the cluster operator excluding/including the core (1s) orbitals of the C atoms, respectively; the separate treatment of the core correlation effects is due to the inadequate treatment of such effects by the current explicitly correlated F12 formalism. The coupled-cluster standard enthalpy of formation at 298.15 K was then estimated

33

as

$$\Delta_f H_{\text{CC}}^{\ominus}(\text{C}_m\text{H}_{2m+2}; 298.15 \text{ K}) = \Delta E_{\text{CC}}(\text{C}_m\text{H}_{2m+2}) - \Delta E_{\text{ZPE}}(\text{C}_m\text{H}_{2m+2})$$
$$+ m \left( \Delta_f H_{\text{exp}}^{\ominus}(\text{C}; 0 \text{ K}) + \Delta H_{\text{exp}}^{\ominus}(\text{C(cr)}; 0 \text{ K} \to 298.15 \text{ K}) \right)$$
$$+ (2m+2) \left( \Delta_f H_{\text{exp}}^{\ominus}(\text{H}; 0 \text{ K}) + \Delta H_{\text{exp}}^{\ominus}(\text{H}_2; 0 \text{ K} \to 298.15 \text{ K}) \right)$$
$$+ \Delta H_{\text{B3LYP/pc-2}}^{\ominus}(\text{C}_m\text{H}_{2m+2}; 0 \text{ K} \to 298.15 \text{ K}), \quad \text{(S17)}$$

where $\Delta E_{\text{ZPE}}(\text{C}_m\text{H}_{2m+2})$ is the zero-point correction to the atomization energy obtained from the B3LYP/pc-2 harmonic vibrational frequencies scaled by 0.985[88]. $\Delta_f H_{\text{exp}}^{\ominus}(\text{X}; 0 \text{ K})$ is the standard enthalpy of formation at 0 K (the experimentally-derived ACtC[89] (version 1.122r) values for carbon (X=C) and hydrogen (H) atoms are 170.028 and 51.633 kcal/mol, respectively), $\Delta H_{\text{exp}}^{\ominus}(\text{X}; 0 \text{ K} \to 298.15 \text{ K})$ is the experimental standard enthalpy difference between 298.15 K and 0 K (the CODATA values for graphite (X=C(cr)) and molecular hydrogen (X=H$_2$) are 0.251 and 2.024 kcal/mol, respectively[90]), and $\Delta H_{\text{B3LYP/pc-2}}^{\ominus}(\text{C}_m\text{H}_{2m+2}; 0 \text{ K} \to 298.15 \text{ K})$ is the standard enthalpy difference between 298.15 K and 0 K for the $n$-alkanes estimated at the B3LYP/pc-2 level of theory using the ideal-gas/rigid-rotor/harmonic-oscillator molecular partition functions. Experimental enthalpies of formation at 298.15 K were taken from Ref. 91.

The remaining errors of the coupled-cluster enthalpies of formation are due to primarily the residual basis set incompleteness errors, the neglect of the accounting for multiple minima on the potential energy surface, the approximate (scaled harmonic) treatment of zero-point energy, the lack of higher-order (post-CCSD(T)) correlation effects, and the neglect of the relativistic effects and post-Born-Oppenheimer effects.

## S5.2 Computational details of the local coupled-cluster computations

All numerical experiments utilizing the DLPNO-CCSD(T)-F12 model of computation were performed using ORCA 4.2[92] on multi-core nodes with dual E5-2683v4 2.1 GHz (Broadwell) processors with a total memory of 512 GB. We demonstrated the near-linear complexity of the DLPNO-CCSD(T)-F12 algorithm by plotting the computational wall timings vs. the system size for $n$-alkanes with model quasi-linear geometries, ranging from $C_{20}H_{42}$ to $C_{120}H_{242}$. The following basis-set triplet (orbital, density fitting, CABS) was used: def2-TZVP, def2-TZVP/C and cc-pVDZ-F12/OptRI. The exchange operator was approximated using the chain of spheres algorithm (COSX) while the Coulomb operator was constructed using the density-fitting procedure from the def2/J basis. All these calculations utilized the `VeryTightSCF` and `TightPNO` input settings of ORCA for higher accuracies. The effective scaling exponent using the $C_{100}$ and $C_{120}$ data points comes out to be around 1.38. The deviation from linearity can be attributed to procedures like the construction of the density-fitted Coulomb operator, which scales at best quadratically with the system size. The near-linear complexity of the DLPNO-CCSD(T)-F12 method has now enabled accurate calculations of the total energies of systems as large as proteins. In this work, we report one such calculation on a model fragment of the nonheme iron-bicarbonate complex in photosystem II (PSII)[93, 94] that plays a key role in photosynthesis. We used the basis-set triplet of {def2-SVP, def2-SVP/RI, `uncontracted` def2-TZVPP} for this simulation with `VeryTightSCF` and `NormalPNO` input settings. Using four computer cores and an intra-node MPI parallel programming model, the total computational time for this energy calculation involving 565 atoms, 2170 electrons and 5420 basis functions, was less than 3 days (70 hours).

# S6 Tensor network calculations

## S6.1 PEPS-DMRG Hubbard calculations

The two-dimensional Hubbard model tensor network calculations were performed with `quimb` [95] using a regular rectangular PEPS [96] ansatz with open boundaries for the $4 \times 4$ lattice at half-filling and the $4 \times 4$, $4 \times 8$, and $4 \times 16$ lattices at 1/8 doping with $U = 8$ and open boundary conditions in both directions. $U(1)$ symmetry was imposed to constrain the number of electrons. Starting from a product state, an initial guess was generated using imaginary time evolution of the PEPS via a Simple Update [97] (SU) scheme. Each iteration of the SU approach comprised a sweep over all of the Trotterized terms in the Hamiltonian, iteratively applying the imaginary time evolution gate to the corresponding nearest-neighboring PEPS site tensors. To ensure convergence, many SU sweeps were performed, with $N_{sweep} \in \{100, 200, 200, 400\}$, for decreasing imaginary time evolution step sizes of $\Delta\tau \in \{0.5, 0.1, 0.05, 0.01\}$.

To further improve upon the ground state, a PEPS-DMRG ground state optimization [98, 99] was used, which mimics the variational minimization done by the MPS-DMRG algorithm by sweeping over all lattice sites and solving localized eigenproblems at each site. To set up the local optimizations, each term in the Hamiltonian is projected into a local basis by performing an approximate boundary contraction [100] of the tensor network with Hamiltonian operators applied to the corresponding local tensors. Reuse of intermediates is used to limit the computational cost scaling for the optimization to $\mathcal{O}\left(L_x^2 L_y^2\right)$. The accuracy of the projected Hamiltonian, and thus the optimization, is controlled by the boundary bond dimension $\chi$, which was set to $\chi = 2D^2$ for all calculations here. This choice makes the total computational cost for given optimization sweep explicitly poly$(D)$. The contraction error for finite $\chi$ is assessed for different square clusters up to size $16 \times 16$ in Fig. S12, which shows that the contraction error is insensitive to the lattice size in this ground-state. Because of the approximate nature of the

projected Hamiltonian, an exact solution of the localized eigenproblem can lead to convergence to local minima or numerical instabilities. To avoid this, only a few iterations of a iterative eigensolver are performed at each site during a sweep; alternatively, one could use a linear combination of the initial guess and exact solution of the localized eigenstates to take small steps in the direction of the global solution. The PEPS results are summarized in Table S4.

To benchmark the 2D tensor network calculations, we performed spin-adapted MPS-based DMRG using BLOCK2 [101]. The exact energy per site and the error are estimated using the same extrapolation method as described in SI section S3.3. The MPS results are shown in Table S5.

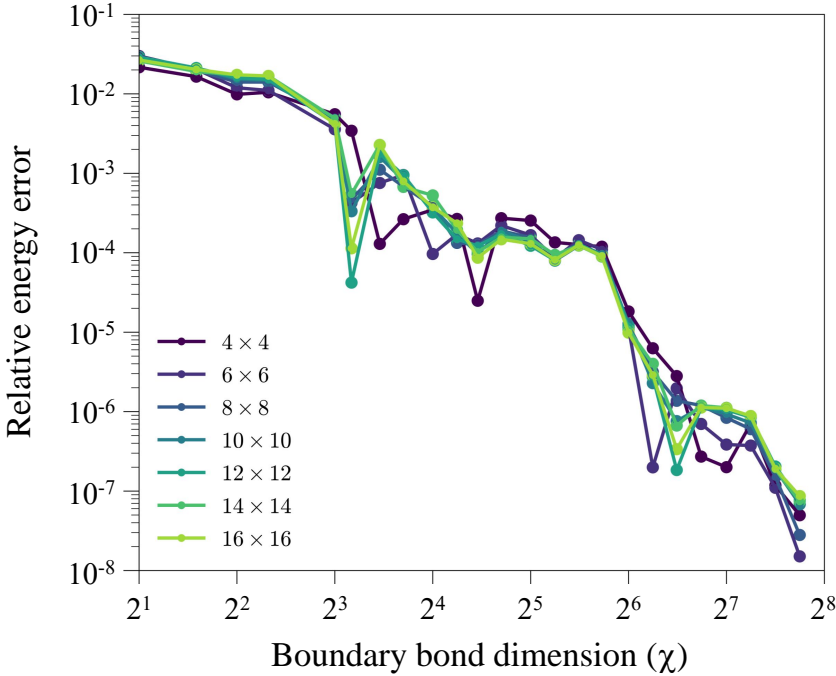

Figure S12: Measuring the contraction error in the PEPS contraction. Relative energy error of the ground-state of the two-dimensional Hubbard model ($t = 1$ and $U = 8$) at $1/8$ doping as a function of the boundary bond dimension ($\chi$) for different lattice sizes ($L$) from $4^2$ to $16^2$. We optimized the energies by the SU scheme with a fixed PEPS bond dimension ($D$) of 8. We used the energies with the largest boundary bond dimension ($\chi = 2^8$) as the reference for the error calculation. We see that the contraction error is insensitive to lattice size in this ground-state.

Table S4: PEPS-SU, PEPS-DMRG, and PEPS-VMC energies per site of the ground state (units of $t$) for the two-dimensional Hubbard models with $t = 1$ and $U = 8$. $D$ represents the maximal virtual bond dimension in the PEPS ansatz. Values in parenthesis denote MC errors at the last digit.

| $D$ | $4 \times 4$, half-filling | | $4 \times 4$, 1/8 doped | | $4 \times 8$, 1/8 doped | | $4 \times 16$, 1/8 doped | | |
|---|---|---|---|---|---|---|---|---|---|
| | SU | DMRG | SU | DMRG | SU | DMRG | SU | DMRG | VMC |
| 3 | | | -0.5256 | -0.5393 | -0.5543 | -0.5727 | -0.5704 | -0.5922 | |
| 4 | -0.4054 | -0.4073 | -0.5576 | -0.5847 | -0.5865 | -0.6095 | -0.5962 | -0.6182 | -0.6182(2) |
| 5 | -0.4105 | -0.4146 | -0.5694 | -0.6013 | -0.5986 | -0.6233 | -0.6091 | -0.6414 | |
| 6 | -0.4154 | -0.4189 | -0.5838 | -0.6089 | -0.6171 | -0.6370 | -0.6313 | -0.6526 | -0.6526(1) |
| 7 | -0.4177 | -0.4230 | | | | | | | |
| 8 | -0.4180 | -0.4238 | -0.5979 | -0.6181 | -0.6285 | -0.6532 | -0.6415 | -0.6624 | -0.6626(2) |
| 10 | -0.4208 | -0.4246 | -0.6056 | -0.6220 | | | | | -0.6688(1) |
| 12 | | | | | | | | | -0.6727(1) |
| 14 | | | | | | | | | -0.6760(1) |

Table S5: MPS-DMRG energies per site of the ground state (units of $t$) for the two-dimensional Hubbard models with $t = 1$ and $U = 8$. $D$ represents the maximal bond dimension in the spin-adapted MPS ansatz.

| $D$ | Discarded weight | Energy per site | $D$ | Discarded weight | Energy per site |
|---|---|---|---|---|---|
| | $4 \times 4$ at half-filling | | | $4 \times 4$ at 1/8 doped | |
| 500 | $3.59 \times 10^{-6}$ | -0.42551504 | 500 | $7.30 \times 10^{-5}$ | -0.63241788 |
| 1000 | $2.61 \times 10^{-7}$ | -0.42552519 | 1000 | $5.40 \times 10^{-6}$ | -0.63260460 |
| 1500 | $5.80 \times 10^{-8}$ | -0.42552583 | 1500 | $1.01 \times 10^{-6}$ | -0.63261660 |
| 2000 | $1.50 \times 10^{-8}$ | -0.42552589 | 2000 | $2.50 \times 10^{-7}$ | -0.63261805 |
| 2500 | $4.12 \times 10^{-9}$ | -0.42552590 | 2500 | $7.55 \times 10^{-8}$ | -0.63261830 |
| $\infty$ | | -0.42552596(1) | $\infty$ | | -0.6326190(1) |
| | $4 \times 8$ at 1/8 doped | | | $4 \times 16$ at 1/8 doped | |
| 500 | $3.66 \times 10^{-4}$ | -0.66601570 | 6000 | $2.27 \times 10^{-5}$ | -0.68515566 |
| 1000 | $1.87 \times 10^{-4}$ | -0.66842805 | 7000 | $1.87 \times 10^{-5}$ | -0.68522094 |
| 1500 | $1.13 \times 10^{-4}$ | -0.66918879 | 8000 | $1.62 \times 10^{-5}$ | -0.68526802 |
| 2000 | $7.59 \times 10^{-5}$ | -0.66952323 | 9000 | $1.41 \times 10^{-5}$ | -0.68530352 |
| 2500 | $5.38 \times 10^{-5}$ | -0.66969776 | $\infty$ | | -0.68555(5) |
| $\infty$ | | -0.67047(15) | | | |

## S6.2   PEPS-VMC Hubbard calculations

In addition to the PEPS-DMRG calculations, we also used the variational Monte Carlo method (VMC) to obtain PEPS energies at bond dimensions $\geq 5$ for the $16 \times 4$ model [102, 103, 104]. This method will be denoted PEPS-VMC. In this approach, the PEPS wavefunction is sampled by Markov Chain Monte Carlo, and the weight of a configuration is computed by contracting a single-layer PEPS tensor network; the accuracy of the contraction is controlled by a boundary bond dimension $\chi$. Both physical observables and energy gradients are evaluated using the importance sampling. In the optimization process of PEPS-VMC, the initial guess is from the aforementioned SU approach, and then the state is further optimized by stochastic gradient optimization. The optimization step length $\delta$ is gradually reduced according to the magnitude of the energy change, from $\delta = 0.005$ to $0.0005$. The optimization is terminated when the energy does not improve (converged up to 3 decimal places) after 50 optimization steps with $\delta = 0.0005$. The cost of each optimization iteration is $O(L_x^2 L_y^2 \chi^2 D^4)$, and we find $\chi = 3D$ works well to produce converged results for all calculations given in Table S4.

The PEPS-VMC method gives consistent results with the PEPS-DMRG method within the statistical error of Monte Carlo sampling. The orange curve shown in the bottom panel of Fig. 4B in the main text ($16 \times 4$ Hubbard model) denotes PEPS-DMRG energies for $D < 5$ and PEPS-VMC energies for $D \geq 5$.

## S6.3   3D Heisenberg model PEPS reference calculations

The 3D anti-ferromagnetic Heisenberg model tensor network simulations were performed with `quimb` [95], using a PEPS [96] ansatz on a cubic lattice with open boundary conditions and Simple Update [97] (SU) style imaginary time evolution. Although more accurate energies could be obtained by variational minimization, we have not performed those calculations in this work due to the large number of tensors to optimize (up to 1000). The energies of the resulting

states are calculated using a basic boundary contraction scheme.

For each cube length $L=\{3, 4, 6, 8, 10\}$, with total number of sites $N=L\times L\times L$, separate ansatz states with virtual bond dimensions $D=2, 3, 4$ were generated. The SU scheme consists of a fixed schedule of 5 sweeps at each imaginary time step $\tau = 10^p$ for $p = \{1.0, 0.9, 0.8, \ldots -3.0\}$ starting from a random PEPS regardless of $D$. Each sweep uses a randomly generated edge coloring of the underlying graph to Trotterize the gates.

The boundary contraction algorithm contracts each face of the cube inwards as a 2D PEPS, eagerly compressing between the tensors as they are contracted into the next plane. No gauging is performed other than performing the SVDs in the reduced basis and ensuring the truncated singular values are absorbed evenly. The actual error values and computational costs are taken using $\chi = \{8, 14, 20\}$ for $D = \{2, 3, 4\}$, respectively, corresponding to $\chi = D^2 + 4$ rounded to the nearest even integer. Note that for SU the cost of computing the energy (or other local quantities) completely dominates the cost of actually performing the optimization. The uncertainty on the error itself is estimated using the standard deviation of energies computed with $\chi = \{10, 12, ..., 20\}$.

The reference energies were computed with ALPS [105], using stochastic series expansion quantum Monte Carlo [106, 107, 108], and were converged to an accuracy of $\approx 10^{-5}$ using a temperature of $T = 0.003J$, where $J$ is the coupling strength.

[56] Zgid, D. & Chan, G. K.-L. Dynamical mean-field theory from a quantum chemical perspective. *The Journal of chemical physics* **134**, 094115 (2011).

[57] Welborn, M., Tsuchimochi, T. & Van Voorhis, T. Bootstrap embedding: An internally consistent fragment-based method. *The Journal of Chemical Physics* **145**, 074102 (2016).

[58] Biroli, G. & Kotliar, G. Reply to "comment on 'cluster methods for strongly correlated electron systems'". *Physical Review B* **71**, 037102 (2005).

[59] Li, Z. & Chan, G. K.-L. Spin-projected matrix product states: Versatile tool for strongly correlated systems. *Journal of chemical theory and computation* **13**, 2681–2695 (2017).

[60] Li, Z., Guo, S., Sun, Q. & Chan, G. K.-L. Electronic landscape of the p-cluster of nitrogenase as revealed through many-electron quantum wavefunction simulations. *Nature chemistry* **11**, 1026–1033 (2019).

[61] Li, Z., Li, J., Dattani, N. S., Umrigar, C. & Chan, G. K.-L. The electronic complexity of the ground-state of the femo cofactor of nitrogenase as relevant to quantum simulations. *The Journal of chemical physics* **150**, 024302 (2019).

[62] Sharma, S., Sivalingam, K., Neese, F. & Chan, G. K.-L. Low-energy spectrum of iron–sulfur clusters directly from many-particle quantum mechanics. *Nature chemistry* **6**, 927–933 (2014).

[63] Peters, J. W. *et al.* Redox-dependent structural changes in the nitrogenase p-cluster. *Biochemistry* **36**, 1181–1187 (1997).

[64] Ohki, Y., Sunada, Y., Honda, M., Katada, M. & Tatsumi, K. Synthesis of the p-cluster



\end{document}